\documentclass[useAMS,usenatbib,authoryear]{mn2e}

\usepackage{graphicx} 
\usepackage{dcolumn}  
\usepackage{bm}       
\usepackage{amssymb,amsmath}  
\usepackage{color}
\usepackage{float}
\usepackage{natbib}
\def\Mpc{\, h^{-1} \, {\rm Mpc}}
\def\kpc{\, h^{-1} \, {\rm kpc}}
\def\Mo{\, h^{-1} \, {\rm M_{\odot}}}

\newcommand{\galacticus}{\textsc{Galacticus}}
\newcommand{\galics}{\textsc{GalICS 2.0}}
\newcommand{\morgana}{\textsc{Morgana}}
\newcommand{\sag}{\textsc{Sag}}

\newcommand{\ysam}{\textsc{ySAM}}
\newcommand{\galform}{\textsc{Galform}}

\newcommand{\dlb}{\textsc{DLB07}}
\newcommand{\lgalaxy}{\textsc{Lgalaxies}}
\newcommand{\sage}{\textsc{Sage}}
\newcommand{\mice}{\textsc{Mice}}
\newcommand{\skibba}{\textsc{SkibbaSUBs}}
\newcommand{\skibbanfw}{\textsc{SkibbaHOD}}

\begin{document}

\title[nIFTY Cosmology: the clustering analysis]{nIFTy Cosmology: the clustering consistency of galaxy formation models}

\author[Arnau Pujol, Ramin A. Skibba, Enrique Gazta\~{n}aga et al.]{\parbox{\textwidth}{Arnau Pujol\thanks{E-mail: arnau.pujol@cea.fr}$^{1,2}$, Ramin A. Skibba$^{3,4}$, Enrique Gazta\~{n}aga$^{1}$, Andrew Benson$^{5}$, \\
Jeremy Blaizot$^{6,7,8}$, Richard Bower$^{9}$, Jorge Carretero$^{10}$, Francisco J. Castander$^{1}$, \\
Andrea Cattaneo$^{11}$, Sofia A. Cora$^{12,13}$, Darren J. Croton$^{14}$, Weiguang Cui$^{15}$, \\ 
Daniel Cunnama$^{16,17}$, Gabriella De Lucia$^{18}$, Julien E. Devriendt$^{19}$, Pascal J. Elahi$^{20}$, \\
Andreea Font$^{21}$, Fabio Fontanot$^{18}$, Juan Garcia-Bellido$^{22,23}$, \\
Ignacio D. Gargiulo$^{12,13}$, Violeta Gonzalez-Perez$^{9,24}$, John Helly$^{9}$, \\
Bruno M. B. Henriques$^{25,26}$, Michaela Hirschmann$^{27}$, Alexander Knebe$^{23,28}$, \\
Jaehyun Lee$^{29}$, Gary A. Mamon$^{27}$, Pierluigi Monaco$^{18,30}$, Julian Onions$^{31}$, \\
Nelson D. Padilla$^{12}$, Frazer R. Pearce$^{31}$, Chris Power$^{32}$, Rachel S. Somerville$^{33}$, \\
Chaichalit Srisawat$^{34}$, Peter A. Thomas$^{34}$, Edouard Tollet$^{11}$, \\
Cristian A. Vega-Mart\'{i}ınez$^{12}$, Sukyoung K. Yi$^{29}$
}\vspace{0.4cm}\\
$^{1}$Institut de Ci\`{e}ncies de l'Espai, IEEC-CSIC, Campus UAB, 08193 Bellaterra, Barcelona, Spain\\
$^{2}$CosmoStat Laboratory, DRF/IRFU/SEDI-Service d'Astrophysique, CEA Saclay, F-91191 Gif-sur-Yvette Cedex, France\\
$^{3}$Department of Physics, Center for Astrophysics and Space Sciences, University of California, 9500 Gilman Drive, San Diego, CA 92093 \\
$^{4}$University of California, Santa Cruz, Science Communication Program, 1156 High Street, Santa Cruz, CA 95064\\
$^{5}$Carnegie Observatories, 813 Santa Barbara Street, Pasadena, CA 91101, USA\\
$^{6}$Universit\'{e} de Lyon, Lyon, F-69003, France\\
$^{7}$Universit\'{e} Lyon 1, Observatoire de Lyon, 9 avenue Charles Andr\`{e}, Saint-Genis Laval, F-69230, France\\
$^{8}$CNRS, UMR 5574, Centre de Recherche Astrophysique de Lyon ; \'{E}cole Normale Sup\'{e}rieure de Lyon, Lyon, F-69007, France\\
$^{9}$Institute for Computational Cosmology, Department of Physics, University of Durham, South Road, Durham, DH1 3LE, UK\\
$^{10}$ Institut de F\'isica d'Altes Energies (IFAE), The Barcelona Institute of Science and Technology, Campus UAB, 08193 Bellaterra (Barcelona) Spain\\
$^{11}$GEPI, Observatoire de Paris, CNRS, 61, Avenue de l’Observatoire 75014, Paris France\\
$^{12}$Instituto de Astrof\'{i}sica de La Plata (CCT La Plata, CONICET, UNLP), Paseo del Bosque s/n, B1900FWA, La Plata, Argentina\\
$^{13}$Facultad de Ciencias Astron\'{o}micas y Geof\'{i}sicas, Universidad Nacional de La Plata, Paseo del Bosque s/n, B1900FWA, La Plata, Argentina\\
$^{14}$Centre for Astrophysics and Supercomputing, Swinburne University of Technology, Hawthorn, Victoria 3122, Australia\\
$^{15}$Departamento de F\'isica Te\'orica, M\'odulo 15, Facultad de Ciencias, Universidad Aut\'onoma de Madrid, 28049 Madrid, Spain\\
$^{16}$South African Astronomical Observatory, PO Box 9, Observatory, Cape Town 7935, South Africa\\
$^{17}$Department of Physics and Astronomy, University of the Western Cape, Cape Town 7535, South Africa\\
$^{18}$INAF - Astronomical Observatory of Trieste, via Tiepolo 11, I-34143 Trieste, Italy\\
$^{19}$Astrophysics, University of Oxford, Denys Wilkinson Building, Keble Road, Oxford, OX1 3RH, UK\\
$^{20}$Sydney Institute for Astronomy, A28, School of Physics, The University of Sydney, NSW 2006, Australia\\
$^{21}$Astrophysics Research Institute, Liverpool John Moores University, IC2, Liverpool Science Park, 146 Brownlow Hill, Liverpool L3 5RF, UK\\
$^{22}$Departamento de F\'isica Te\'{o}rica, M\'{o}dulo 15, Facultad de Ciencias, Universidad Aut\'{o}noma de Madrid, 28049 Madrid, Spain\\
$^{23}$Instituto de F\'{i}sica Te\'{o}rica, Universidad Aut\'{o}noma de Madrid (IFT-UAM/CSIC), 28049 Madrid, Spain\\
$^{24}$Institute of Cosmology \& Gravitation, University of Portsmouth, Dennis Sciama Building, Portsmouth PO1 3FX, UK\\
$^{25}$Max-Planck-Institut f\"{u}r Astrophysik, Karl-Schwarzschild-Str. 1, 85741 Garching b. M\"{u}nchen, Germany\\
$^{26}$Institute for Astronomy, ETH Zurich, CH-8093 Zurich, Switzerland\\
$^{27}$Institut d'Astrophysique de Paris (UMR 7095: CNRS \& UPMC), 98 bis Bd Arago, F-75014 Paris, France\\
$^{28}$Astro-UAM, UAM, Unidad Asociada CSIC\\
$^{29}$Department of Astronomy and Yonsei University Observatory, Yonsei University, Seoul 120-749, Republic of Korea\\
$^{30}$Dipartimento di Fisica, Universit\`{a} di Trieste, via Tiepolo 11, 34143 Trieste, Italy\\
$^{31}$School of Physics \& Astronomy, University of Nottingham, Nottingham NG7 2RD, UK\\
$^{32}$International Centre for Radio Astronomy Research, University of Western Australia, 35 Stirling Highway, Crawley, \\Western Australia 6009, Australia\\
$^{33}$Department of Physics and Astronomy, Rutgers University, 136 Frelinghuysen Road, Piscataway, NJ 08854, USA\\
$^{34}$Department of Physics \& Astronomy, University of Sussex, Brighton, BN1 9QH, UK\\
}
\date{Accepted xxxx. Received xxx}

\pagerange{\pageref{firstpage}--\pageref{lastpage}} \pubyear{2017}

\maketitle

\clearpage

\label{firstpage}

\begin{abstract}

We present a clustering comparison of 12 galaxy formation models (including Semi-Analytic Models (SAMs) and Halo Occupation Distribution (HOD) models) all run on halo catalogues and merger trees extracted from a single $\Lambda$CDM N-body simulation. We compare the results of the measurements of the mean halo occupation numbers, the radial distribution of galaxies in haloes and the 2-Point Correlation Functions (2PCF). We also study the implications of the different treatments of orphan (galaxies not assigned to any dark matter subhalo) and non-orphan galaxies in these measurements. 
Our main result is that the galaxy formation models generally agree in their clustering predictions but they disagree significantly between HOD and SAMs for the orphan satellites.
Although there is a very good agreement between the models on the 2PCF of central galaxies, the scatter between the models when orphan satellites are included can be larger than a factor of $2$ for scales smaller than $1\Mpc$. We also show that galaxy formation models that do not include orphan satellite galaxies have a significantly lower 2PCF on small scales, consistent with previous studies. 
Finally, we show that the 2PCF of orphan satellites is remarkably different between SAMs and HOD models. Orphan satellites in SAMs present a higher clustering than in HOD models because they tend to occupy more massive haloes.
We conclude that orphan satellites have an important role on galaxy clustering and they are the main cause of the differences in the clustering between HOD models and SAMs.

\end{abstract}


\maketitle

\section{Introduction}

In $\Lambda$CDM cosmology, gravitational evolution causes dark matter to cluster around peaks of the initial density field and to collapse into virialized objects (i.e., dark matter haloes). Structures form hierarchically, such that smaller haloes merge to form larger and more massive haloes.  All galaxies are thought to form as a result of gas cooling at the center of the potential well of dark matter haloes.  When a halo and its `central' galaxy are accreted by a larger halo, it becomes a subhalo and its galaxy becomes a `satellite' galaxy. However, due to tidal stripping and the gravitational interaction of subhaloes with their environment (other subhaloes, the gravitational potencial of the halo centre, etc.), sometimes can be distrupted and the galaxy, if it survives, becomes an `orphan' galaxy. 
In addition to mergers, haloes also grow by smooth accretion and galaxies grow by \textit{in situ} star formation when fuel (i.e., cold gas) is available \citep{Cooray2002,Sheth2002,vandenBosch2002,Gill2004b,Delucia2004,Vandenbosch2005,Diemand2007,Giocoli2008}.

In this paradigm of hierarchical structure formation, there is a correlation between halo formation, their abundances and the surrounding large-scale structure where more massive haloes tend to reside \citep{Mo1996,Sheth2002}.  Most galaxy formation models implicitly assume that the properties of a galaxy are determined primarily by the mass and formation history of the dark matter halo within which it formed \citep{White1978,White1991,Cole1991,Lacey1991,Baugh1999,Benson2001}. Thus, the correlation between halo properties and environment (matter density, substructure, etc.) induces a correlation between galaxy properties and environment.

There are multiple statistical quantities used to study large-scale structure, and here we focus on the two-point correlation function, the radial distribution and the mean occupation number of galaxies. Clustering studies have shown that a variety of galaxy properties (such as luminosity, color, stellar mass, star formation rate and morphology) are dependent on the environment and halo properties across a wide range of scales. Galaxy formation models in simulations are crucial to study the connection between galaxies and haloes, and hence it is important to understand the consistency or differences between different galaxy formation models.

Galaxy formation is a complex, nonlinear process, driven by the interplay of many different physical mechanisms \citep[e.g.][]{Benson2010}. The goal of galaxy formation models is to estimate the statistical properties of the galaxy population given some set of assumptions and thereby to better understand the physical processes involved. One fruitful approach has been to utilize Semi-Analytic Models (SAMs) of galaxy formation \citep[e.g.][]{Cole2000,Hatton2003,Cattaneo2006,Cora2006,Croton2006,Baugh2006,DeLucia2007,Monaco2007,LoFaro2009,Benson2012,Lee2013,Henriques2013,Baugh2013,GonzalezPerez2014,Gargiulo2015}, in which a statistical estimate of the distribution of dark matter haloes and their merger history---either coming from cosmological simulations or extended Press-Schechter/Lagrangian methods---is combined with simplified yet physically motivated prescriptions of processes such as star formation, gas cooling, feedback from supernovae and active galactic nuclei (AGN), etc. that allows one estimate the distribution of galaxy properties. New models are starting now to also use observations of galaxy clustering to constrain their parameters \citep{vanDaalen2016}.

An alternative approach to SAMs are (analytic) dark matter halo occupation models, which determine the halo occupation of galaxies based on the properties of their parent halo. Usually observations of clustering are used to constrain this occupation. This approach is used to study the link between galaxy formation and halo assembly (see \citealt{Cooray2002}; \citealt{Mo2010} for a review). Halo models of galaxy abundances and clustering generally consist of Halo Occupation Distribution (or conditional luminosity functions) \citep[HOD; e.g.][]{Seljak2000,Scoccimarro2001,Berlind2002,Cooray2002,Yang2003,Kravtsov2004,Cooray2006,Guo2016b} and (sub)halo abundance matching (HAM or SHAMs; \citealt{Vale2006}; \citealt{Conroy2006}; \citealt{Hearin2013}; \citealt{Reddick2013}; \citealt{Guo2016b}). Such models are useful for exploring the relations between galaxy formation and dark matter halo assembly in the context of the large-scale structure of the Universe.

Subhaloes closer to the halo centre tend to accrete earlier on \citep{Gao2004}, and therefore tidal stripping (a process which is in part numerical but also physical) has more time to act on these subhaloes. Hence, they are more frequently disrupted and this is the reason why subhaloes are anti-biased with respect to the dark matter \citep[see e.g. ][]{Ghigna2000,Diemand2004,Pujol2014b} in these regions.
In simulations, lack of mass resolution causes the disappearance of a subhalo, causing the galaxy to become an \emph{orphan}. Sometimes the halo finder will merge a subhalo with its parent halo, but the subhalo can reappear when its member particles bounce out of the halo. There are different ways to follow the positions of these orphan galaxies, and these can lead to different clustering of galaxies, especially on small scales  \citep{Gao2004,Wang2006,Guo2011,Budzynski2012,Lee2014}. 
In \cite{Gao2004},
they used high-resolution  resimulations of  galaxy clusters and analysed
the radial density profiles of both  subhaloes and  galaxies from a SAM. This study showed that by including orphan galaxies the radial
density distribution was very close to that  of the dark matter, as inferred in
the observational data. They also argued that increasing the
resolution would not improve the situation. This result has been confirmed in 
 \cite{Guo2011}, who showed that orphans are still dominating the
central regions of galaxy clusters when increasing the resolution of the simulations.  They showed that by tracking the position of the most bound particle at the time of disruption convergence between simulations of different resolution was achieved. In \cite{Wang2006}, an HOD approach was used, but using the number and positions of
galaxies from a SAM, and they showed that orphan galaxies  are
needed  to reproduce  the  clustering  signal at small scales, also confirmed by recent studies \citep{Budzynski2012}. Finally, \cite{Kang2012}, \cite{Guo2013} and \cite{Henriques2013} showed that cosmology, within the current precision, has no impact in the clustering when compared to the differences from galaxy formation physics, even on large scales.

 
The trajectory and lifetime of orphan galaxies can be determined from different approaches in SAMs. On one side, some SAMs immediately merge galaxies with the central galaxy when the subhalo is lost, and then they have no orphan satellites by construction. Other SAMs define an analytical orbit for the orphan galaxies according to the position and velocity of the galaxies when they became orphan. The radius of the orbit is then continuously decreased until it merges with the central galaxy. Finally, other SAMs define the position and velocity of orphan galaxies directly from the dark matter particle that was the most gravitationally bound from the disrupted subhalo. And other SAMs \citep{Guo2011} use a combination of both the analytical orbits with the dark matter particle trajectories. All these different treatments of orphan galaxies can have consequencies on the abundance and distributions of such galaxies, especially at small scales.

SAM and HOD have important differences on the treatment of orphan galaxies. First of all, while SAMs make use of the merger trees to derive the initial trajectories of orphan galaxies, the HOD models define the galaxy distribution from the present distribution of haloes, without using information from their evolution. Moreover, classical HODs do not account for the presence of substructures - they are built on top of dark matter haloes and the population of satellites is just distributed according to an NFW model.  

Given the variety of galaxy formation models that are used in simulations nowadays, it is important to study the differences that arise from the different treatment of galaxy formation physics in each model. In fact, many efforts have been done comparing different galaxy formation models and their physical prescriptions \citep{Somerville1999,Fontanot2009,Kimm2009,Contreras2013,Delucia2011,Fontanot2011,Fontanot2012,Kang2014,Somerville2015,Guo2016b}. This study focuses on the differences in the galaxy clustering for a large variety of models run on the same simulation and with the same merger tree. 

In an attempt to put together a large representation of the models from the literature in an extensive comparison study, \cite{Knebe2015} (K15 hereafter) presented 14 models (12 SAMs and 2 HOD models) using the same simulation input (halo catalogues and merger trees) and analyzed the consistency between the models looking at the stellar mass function, the star formation, stellar-to-halo mass relations, stellar mass fractions or abundance of galaxies per halo. 
The present paper is a complement of K15, where we study the consistency between several galaxy formation models on the clustering and the distribution of galaxies in haloes. We analyze the consequences of the differences between the models on the distribution of the galaxies inside and outside haloes. We do this by comparing the Two Point Correlation Function of galaxies, the halo occupation number and the radial distribution of galaxies in haloes. We also analyse orphan satellites separately in order to focus on the consequences from the different treatments of orphan satellites between HOD models and SAMs. 

This paper is organized as follows. In the next section, Section~\ref{sec:simulation}, we describe the dark matter halo simulation and the orphan treatments of the galaxy formation models. Then we describe our methodology in Section~\ref{sec:methodology}. We present our results, including comparisons of halo occupation numbers, radial distribution of galaxies in haloes and galaxy clustering in Section~\ref{sec:results}. Finally, we end by summarizing and discussing our results in Section~\ref{sec:conclusions}.

\section{Simulation data}\label{sec:simulation}

For this study we use a dark matter halo catalogue generated from a \textsc{Gadget-3} N-body simulation (\citealt{Springel2005b}) of a $62.5 \Mpc$ side box. We use $270^3$ particles with a particle mass resolution of $9.31 \times 10^8 \Mo$, producing an output of $62$ snapshots. From each snapshot we generate a halo catalogue using the \textsc{SUBFIND} (\citealt{Springel2001}) code, that generates haloes and subhaloes from dark matter overdensities. We used the code \textsc{MergerTree} to generate the merger trees of the haloes \footnote{MergerTree forms part of the \textsc{AHF} package (\citealt{Knollmann2009})}.

From the simulation we obtained several mass definitions for the haloes that can be used for the galaxy formation models. The mass definitions used are detailed in Appendix \ref{sec:mass_comp}. 
Some properties of the galaxy formation models can be sensitive to the mass definition and to the galaxy formation models. We discuss the mass definition criteria in Section \ref{sec:methodology}. 

We use several galaxy formation models together with this dark matter only simulation for the comparison analysis. Some of them are SAMs of galaxy formation, while others are based on the HOD model. We refer to K15 for a detailed description of these models and some comparisons between them.  In this section we enumerate the models, their acronyms and references in Table \ref{tab:models}, and briefly describe the treatment and merging of the orphan satellites of each, since this is one of the most relevant aspects for this comparison analysis. 

All the models were originally calibrated to reproduce a given set of observations. However, each model uses different observational data and simulated cosmologies to calibrate its parameters, as stated in the corresponding papers describing the models. It is worth stressing that our strategy forces all models to the same underlying merger tree, therefore we do not expect the original calibrations to be optimal. We have seen in K15 that this leads to model-to-model variations larger than if they were all calibrated for this particular simulation. Nonetheless, we are interested in the general agreement between the different galaxy formation models. 

\begin{table*}
\begin{center}
\begin{tabular}{c c c c l}
\hline
Model & Type & Orphans & Orphan positions & Reference \\
\hline
DLB07 & SAM & YES & NO & \citealt{DeLucia2007} \\
Galacticus & SAM  & YES & NO & \citealt{Benson2012} \\
Galform & SAM  & YES & NO & \citealt{GonzalezPerez2014} \\
GalICS 2.0 & SAM  & NO & - & \citealt{Hatton2003,Cattaneo2006,Cattaneo2017} \\
LGALAXIES & SAM  & YES & YES & \citealt{Henriques2013} \\
MICE & HOD  & YES & YES & \citealt{Carretero2015} \\
MORGANA & SAM  & YES & NO & \citealt{Monaco2007,LoFaro2009} \\
SAG & SAM  & YES & YES & \citealt{Cora2006,Gargiulo2015} \\
SAGE & SAM  & NO & - & \citealt{Croton2016} \\
SkibbaSUBs & HOD  & YES & YES & \citealt{Skibba2006, Skibba2009} \\
SkibbaHOD & HOD  & YES & YES & \citealt{Skibba2006, Skibba2009} \\
ySAM & SAM  & YES & YES & \citealt{Lee2013} \\
\hline
\end{tabular}
\caption{This table shows the list of galaxy formation models used in this paper. The first column shows the acronyms used for each of the models. The second column specifies whether the model is a SAM or HOD. The third column specifies if the model has implemented a treatment of orphan satellites. The fourth column specifies if the model has calculated the positions or orbits of the orphan satellites for this work. Finally, a list of the references is shown in the last column. }
\label{tab:models}
\end{center}
\end{table*}

\subsection{Treatment of orphan satellites}\label{sec:orphan_treatments}

The treatment of the orphan satellites (satellites with no associated dark matter subhalo) has a direct impact on galaxy clustering. In this section we give a brief overview on how models deal with the orphan  population (if any) and we refer the interested reader to K15 for more details on the modeling of other physical processes.

\subsubsection{\dlb, \galacticus\ and \galform\ (SAM)}

In these models, when a subhalo disappears (it is stripped below the resolution of the parent simulation), a merger time is assigned to its galaxies according to some variations of the Chandrasekar formula and galaxies are merged when this time is over. These galaxies are assumed to continue orbiting within their parent halo until dynamical friction causes it to merge with the central galaxy. Positions and velocities of orphan satellites are assumed to be traced by those of the most bound particles of substructures at the last time they were identified. This information was not provided for the simulation used in this study. Therefore, the positions of orphan satellites in these models cannot be used for the clustering analysis presented here.

\subsubsection{\galics\  (SAM)}

 In this model, the effects of the merging timescale are degenerate with those of supernova and AGN feedback and the shock-heating scale. Then, the same effects in the stellar mass function due to the contribution of orphan satellites can also be obtained without them by  lowering the efficiency of supernova feedback or the shock heating mass. This model contains a free parameter that sets the dynamical friction efficiency, producing orphan satellites only when this parameter is larger than $0$ (otherwise galaxies always merge when haloes and subhaloes merge). As a good fit to observations (not incuding clustering observations) was obtained without the need of orphan satellites, we decided to use the simplest solution and set the dynamical friction efficiency parameter to $0$. Because of this, \galics\ has no orphans. 

\subsubsection{\lgalaxy\  (SAM)}

In the original form of \lgalaxy\ the positions of orphans are followed by tracking the most-bound particle of their host dark matter halo just before it was tidally disrupted. The orphan satellite is then placed not at the current position of the particle with which it is identified, but at a position whose (vector) offset from the central galaxy is reduced from that of the particle by a factor of $(1-\delta t/t_{friction})$ where $\delta t$ is the time since the dynamical friction clock was started. This time dependence is based on a simple model for a satellite with ``isothermal" density structure spiralling to the centre of an isothermal host on a circular orbit \citep{Guo2013,Henriques2013}. 

Since the dynamical information of most-bound particles is not available for the current simulation, for this work \lgalaxy\ simply decays the positions of orphans from their value at the time they become orphans. Instead of $(1-\delta t/t_{friction})$, a factor of $2 \times
\sqrt{1-\delta t/t_{friction}}$ is used in order to obtain satellite profiles and small scale clustering that roughly resemble those from the default model.

\subsubsection{\mice\ (HOD)}

The galaxy population in haloes is determined from the halo mass, independently of their substructure. In the original implementation of the model, the luminosity function and the colour-magnitude diagrams are determined from observations \citep{Blanton2005b}. Then, the galaxies are split into centrals and satellites. Using a modified NFW profile for the satellite distribution inside haloes, the occupation of galaxies as a function of halo mass is calibrated in order to reproduce the 2-Point Correlation Function (2PCF) of galaxies from observations. The modification of the NFW profile corresponds to a slightly steeper distribution that improves the clustering consistency with observations \citep{Zehavi2011}. 

For this particular project, once the number of satellite galaxies in a halo is set, each satellite galaxy is assigned to a different subhalo. When there are more satellites than subhaloes, the excess of satellites are considered orphans and populate the halo according to a modified NFW profile. This is not the approach used in the original implementation of the model, where all the satellite galaxies are distributed with the modified NFW profile. In this case, we will be able to study the clustering of galaxies that follow subhaloes (the non-orphan satellites) and those that are consistent with the original implementation of the HOD model (the orphan satellites). 

Although originally this model was implemented and calibrated in the MICE simulation \citep{Crocce2015,Fosalba2015,Fosalba2015b} to make it consistent with clustering observations, in this simulation we used the same parameters obtained from the calibration in the MICE simulation, and hence the model is not necessarily reproducing the clustering observations in this study. In order to reproduce clustering observations, we would need to recalibrate the parameters of the model in this simulation and cosmology (the original simulation used the parameters $\Omega_m = 0.25$, $\Omega_\Lambda = 0.75$ and $\sigma_8 = 0.8$, while this simulation has $\Omega_m = 0.272$, $\Omega_\Lambda = 0.728$ and $\sigma_8 = 0.807$, to mention some parameters). 

\subsubsection{\morgana\ (SAM)}

This model has been originally designed to work with merger trees
generated by the Lagrangian code \textsc{Pinocchio} \citep{Monaco2002}, and hence some adjustments have been needed in order to use it
interface with \textsc{SUBFIND} based merger trees.

As \morgana\ does not explicitely follow the evolution of substructures,
only central galaxies are linked to a dark matter structure. Whenever a dark matter halo
becomes a substructure, its galaxies become satellites and each of them
receive a merging time (computed from the \cite{Taffoni2003}
prescriptions), which is defined independently from substructure
evolution. As these merger times are estimated statistically, the
merger of a satellite galaxy with the central object is decoupled from
its parent substructure survival, i.e. satellite galaxies may merge
before their host substructure is lost  (giving rise to a population
of substructures whose galaxies have already disappeared; the other models assign a residual merger time when the subhalos is lost, assuring that no merger takes place before the subhalo disappears) or, vice versa,
after it.  In both cases they are considered as orphans, and placed at
the centre of the host main halo (meaning that we do not track the trajectories of the orphans). As this assignment of the position
of the orphans is not physical when describing small scale clustering,
we exclude them for the analysis of this model. The other non-central galaxies are considered non-orphan satellites for this analysis (differently than in K15, where all satellite galaxies of this model are considered orphans), and their position is then defined from their subhaloes.

Moreover, merging times computed from the \cite{Taffoni2003}
prescriptions are typically shorter than those estimated from N-Body
simulations \citep{Delucia2010}. Both effects have important
implications on the satellite number density, showing a lower halo occupation number as we discuss later.

\subsubsection{\sag\ (SAM)}

When the subhaloes are no longer identified due to the mass loss form the merging with a larger structure, their galaxies become orphans. The trajectory of orphan satellites is calculated to be a circular orbit with a velocity determined by the virial velocity of the host subhalo and initially located at a halocentric distance given by the virial radius of the subhalo. The decaying radial distance is estimated from the dynamical friction, with position and velocity components randomly generated. The orphan satellites finally merge with the central galaxy of the substructure in which they reside according to the dynamical friction time-scale \citep{Binney1987b}. Because of this time-scale, orphan satellites can be found inside the biggest substructure of the halo (where the central galaxy resides) or inside the substructure of another (satellite) galaxy.

\subsubsection{\sage\ (SAM)}

When a halo/central galaxy system is captured by something larger to become a subhalo/satellite galaxy, the \emph{expected} average merger time of the system is calculated using the \cite{Binney1987b} dynamical friction formula. The subhalo/satellite is then tracked with time until its subhalo-to-baryonic mass ratio falls below a critical threshold, taken as 1.0. At this point the current survival time as a subhalo/satellite is compared to the expected merger time calculated at infall. If the subhalo/satellite has survived longer than average we say it is more resistant to disruption and the satellite is merged with the central in the usual way. Otherwise the satellite is disrupted and its stars are added to a new intra-``cluster'' mass reservoir. As a consequence, SAGE does not produce an orphan galaxy population, since the decision about (and implementation of) the ultimate fate of a satellite is always made before (or when) its subhalo is lost in the merger tree.

This model is an update of \cite{Croton2006}, and the suppression of orphan galaxies and the satellite treatment are some of the changes. Satellite galaxies in \cite{Croton2006} were found to be too red, mainly because of the instantaneous hot gas stripping that was causing a premature supression of star formation. In \sage\ satellite galaxies are treated more like central galaxies, in the sense that hot gas stripping now happens in proportion to the subhalo mass stripping. The lack of orphans makes the model to be resolution dependent, since the population of satellite galaxies depends on the resolution and detection of subhaloes in the simulations. However, most modern cosmological simulations have sufficiently high resolution to recover the galaxy population down to the limit of that typically probed by current surveys \citep{Croton2016}. 

\subsubsection{\skibba\ (HOD)}

As in \mice, the occupation of galaxies in haloes are determined as a function of the halo mass, independently of the substructure, and calibrated to recover luminosity function, colour magnitude diagrams and the clustering 2PCF \citep{Blanton2003,Zehavi2005}. 
Satellite galaxies are distributed in subhaloes, and the exceeding galaxies are considered orphan satellites and populate the haloes according to the NFW profile. For an additional comparison in this study, we also constructed a catalogue where all the satellite galaxies follow the NFW profile (independently of the substruture of the haloes), that we call \skibbanfw. The difference between \skibba\ and \skibbanfw\ is that the satellite galaxies have a different density profile. In \skibba\ the satellite galaxies follow the subhaloes (except the orphan satellites) while in \skibbanfw\ all satellite galaxies follow an NFW profile. This is useful to understand the importance of the different treatments of the satellite distribution and how they affect clustering. In this study \skibbanfw\ will be treated as a reference for an NFW-based model. 

As in \mice, the parameters have been calibrated in another simulation and cosmology, and hence the model does not necessarily reproduce clustering observations even they were fit to do it in the original simulation. 

\subsubsection{\ysam\  (SAM)}

All subhaloes are tracked even after the halo finder loses them in the central dense region of a main halo, and populated with a resident galaxy. In these cases, galaxies only merge when they are closer to the centre of the halo than $0.1R_{vir}$. If a substructure disappears before reaching the central region of its host halo, the galaxy is considered orphan and \ysam\ calculates its mass \citep{Battin1987} and orbit \citep{Binney1987b} analytically until approaching the very central regions. This has a large impact on the lifetime of subhaloes and galaxy merging timescale \citep{Yi2013}.

\subsection{Orphan fraction}

\begin{figure}
\begin{centering}
\includegraphics[scale = .58]{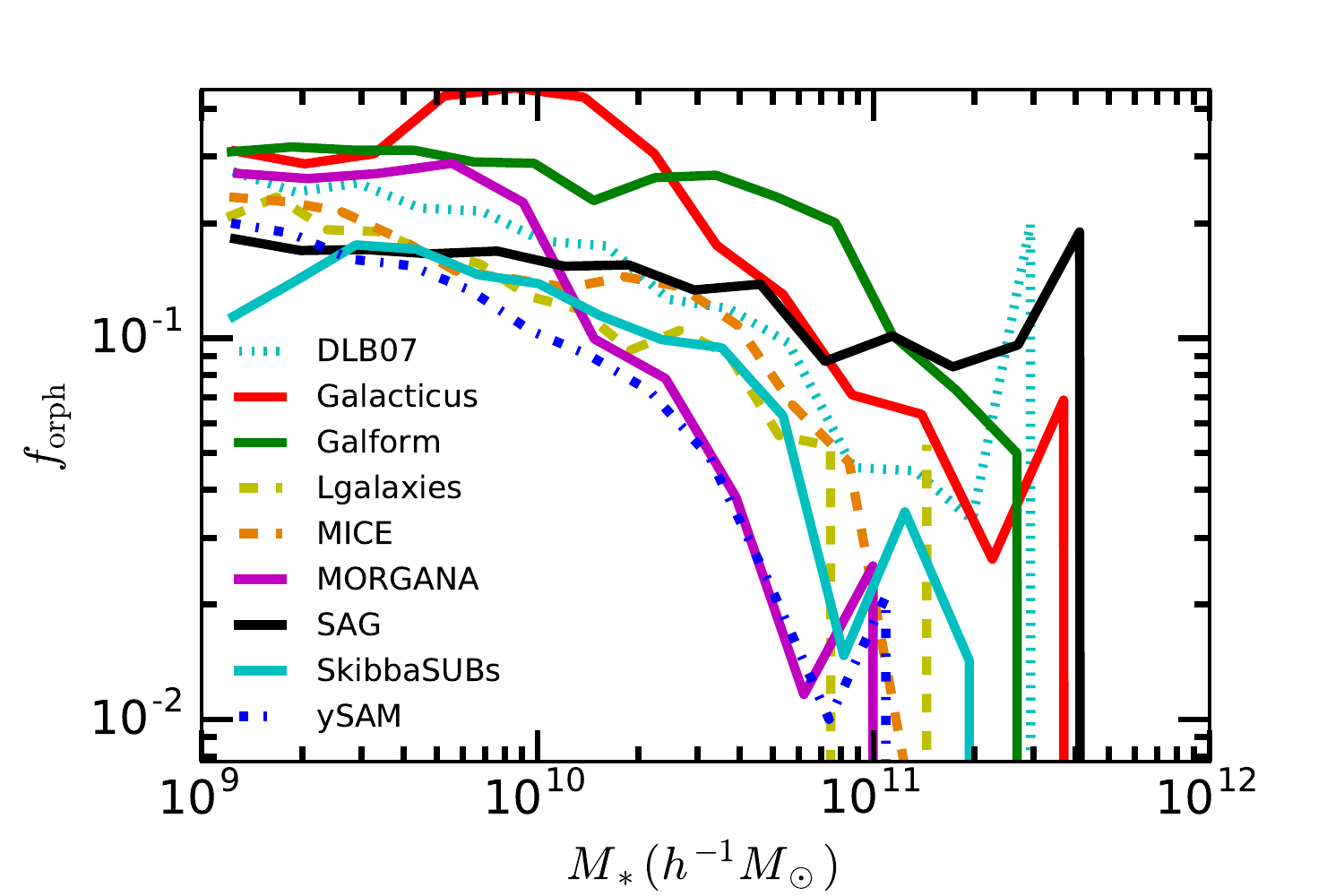}
\caption[Orphan fraction as a function of stellar mass]{Comparison of the orphan fraction (with respect to all the galaxies) of the different galaxy formation models as a function of the stellar mass. Each line corresponds to a different model. \sage\ and \galics\
 did not consider orphans in this work.}
\label{fig:orph_frac}\par\end{centering}
\end{figure}

In Fig. \ref{fig:orph_frac} we show the orphan fraction as a function of stellar mass for the galaxy formation models that have orphan satellites. This figure is similar to Fig.12 in K15 where the same fraction has been plotted, but as a function of halo mass. The orphan fraction is defined as:

\begin{equation}
f_{\mbox{orph}} (M_*) = \frac{N_{\mbox{orph}}(M_*)}{N(M_*)},
\end{equation}
where $N_{\mbox{orph}}(M_*)$ is the number of orphan satellites in the catalogue with stellar mass $M_*$ and $N(M_*)$ is the total number of galaxies with the same stellar mass. We see that the fraction decreases with mass. 
This trend is expected since small subhaloes are more easily affected by tidal stripping, so orphan satellites tend to originate from small subhaloes.  As the mass of the orphan satellites is strongly related to the mass of the subhaloes at the time of accretion, their masses then tend to be small. 
Because of this, orphan satellites are more important at small masses than at large masses, and this implies that the role of orphan satellites on galaxy clustering will mostly be important for low mass thresholds \citep{Gao2004,Wang2006,Guo2011,Budzynski2012,Lee2014}. The large orphan fractions of \dlb, \galacticus\ and \sag\ at $M_* > 3\times 10^{11}\Mo$ are not significant due to the low number of galaxies with these masses, that makes very few orphan satellites (less than 5)  represent a large fraction.   

We can see that the fraction of orphan satellites depends strongly on the different galaxy formation model, with the scatter between the models being large. The models that show the lowest orphan fractions (for high masses) are \ysam\ and \morgana, as expected from their treatment of orphan satellites. On one hand, \ysam\ tracks the galaxies in the substructures even after its mass has decreased below the resolution level. These galaxies are still considered non-orphan galaxies, and because of this it is more difficult for a galaxy to become orphan in this model. On the other hand, in \morgana\ the galaxy merger times are shorter than those estimated from N-body simulations, which means that orphan satellites merge more quickly with central galaxies. Therefore there are fewer orphans for this model. Interestingly, \galacticus\ and \galform\ show the highest orphan fractions. Several studies \citep{Contreras2013,Campbell2015,Simha2016} show that the analytical equation used in \galform\ allows galaxies to orbit around the central galaxy for longer than other approximations, also causing a more centrally concentrated distribution of satellite galaxies. Note that the models that track the positions of the orphan satellites, and hence the models that we will use to measure the clustering of orphan satellites, are the ones that present the lowest orphan fraction at small masses. This means that the results of the distribution of orphan satellites that we show in this study might have a stronger impact on galaxy clustering for the rest of the models.

\section{Methodology}\label{sec:methodology}

\begin{figure*}
\begin{centering}
\includegraphics[scale = .43]{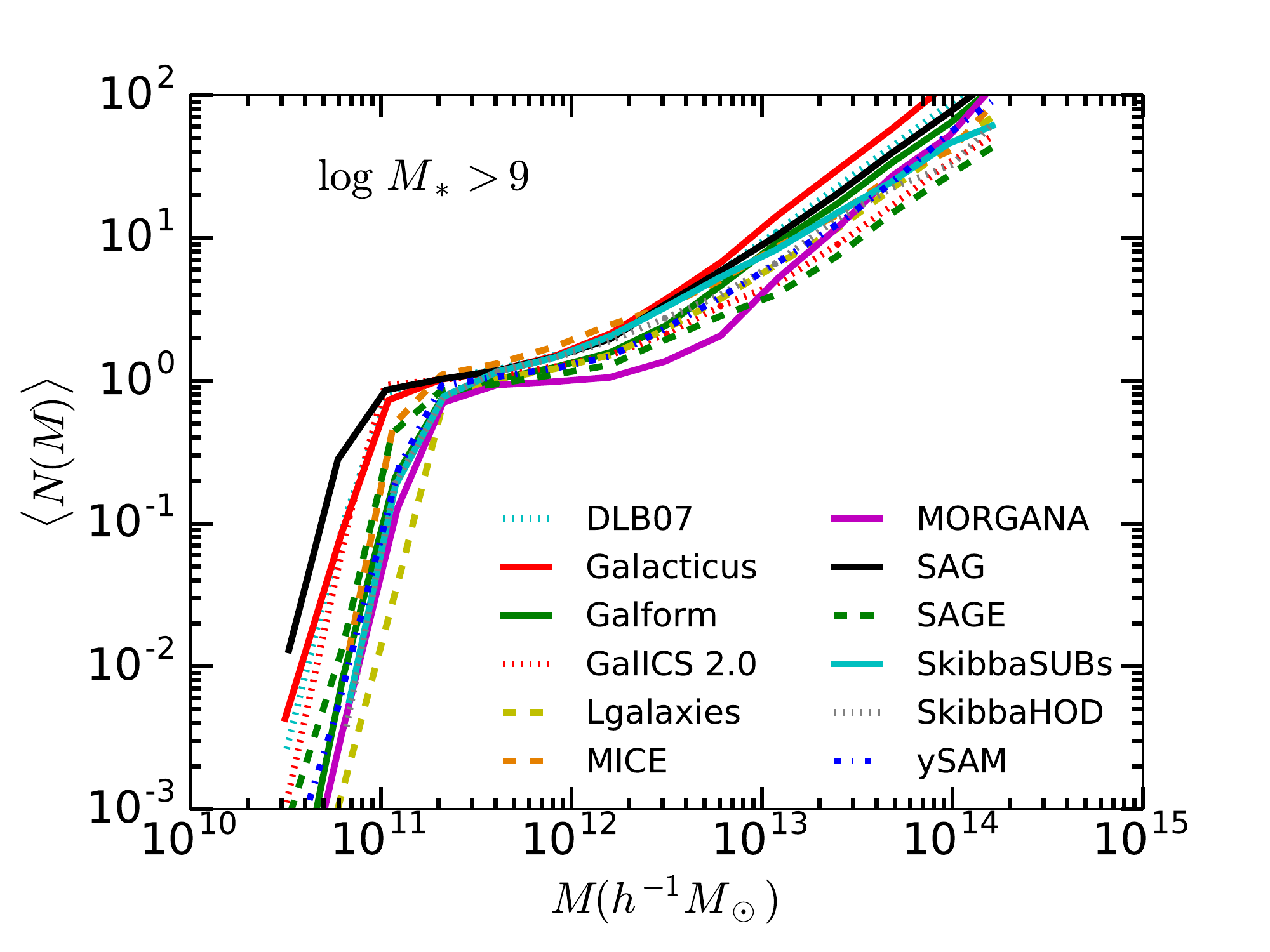}
\includegraphics[scale = .43]{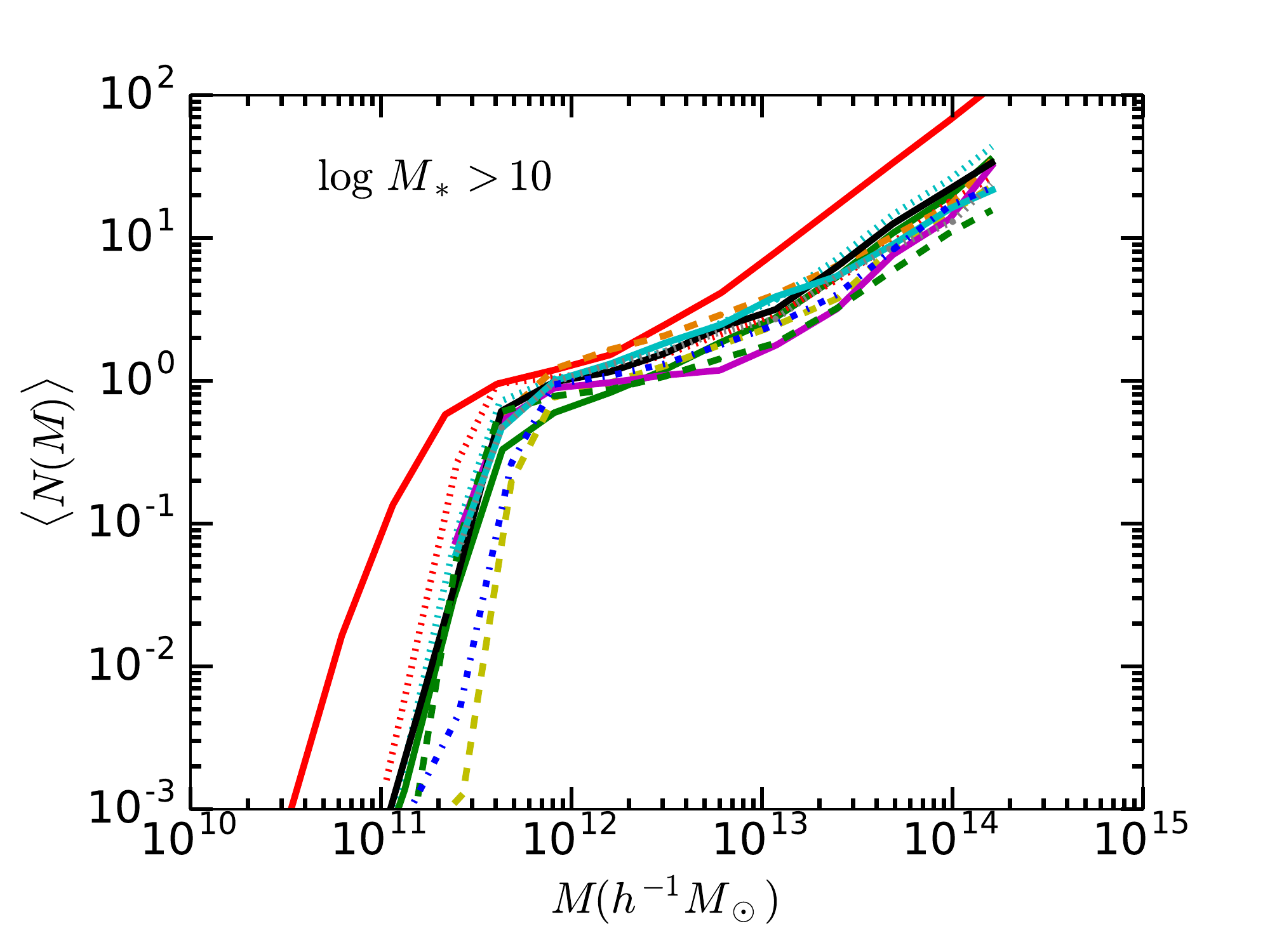}
\caption[Mean halo occupation number comparison of the different models]{Mean halo occupation number as a function of halo mass for the different models using all the galaxy types. Left panels show galaxies with $M_* > 10^9 \Mo$, while the right shows galaxies with $M_* > 10^{10} \Mo$. }
\label{fig:hod_comp}\par\end{centering}
\end{figure*}

To study the clustering between the models, we will use the 2PCF, which describes the excess of probability $dP$ over a random distribution of finding pairs of galaxies at a given separation $r$:

\begin{equation}
dP = n(1 + \xi(r))dV,
\end{equation}
where $\xi(r)$ represents the 2PCF at a separation $r$ and $n$ is the number density of galaxies. There are several estimators of the 2PCF \citep{Kerscher2000,Coil2013}, but for our study we use the estimator described by the following formula: 
\begin{equation}
\xi(r) = \frac{DD(r)}{RR(r)} -1,
\end{equation}
where $\xi(r)$ is the 2PCF as a function of scale, $DD(r)$ is the number of data pairs separated a distance $r$ between them, and $RR(r)$ is the number of random pairs at the same distance. $DD(r)$ and $RR(r)$ are normalized by $n_D(n_D - 1)$ and $n_R(n_R - 1)$ respectively, where $n_D$ and $n_R$ are the numbers of data and random points used. This estimator is equivalent to \cite{Landy1993} when the random sample is large enough, as it is here, where we use $n_R = 10^6$. Due to the size of the simulations, we calculate $\xi(r)$ up to $R=6 \Mpc$, since the measurement becomes noisy for larger scales. Due to the resolution, the minimum scale for the study of $\xi(r)$ is $R = 300 \kpc$ \citep{Guo2011}.

To calculate the errors of the 2PCF we use the Jack-Knife method \citep{Norberg2009}. We divide the simulation box into $64$ cubic subvolumes, and we measure the 2PCF $64$ times excluding each time one of the subsamples. We obtain the error from these measurements using the unbiased standard deviation according to the following formula:

\begin{equation}
 \Delta \xi(r) = \sqrt{ \left(\frac{N_{JK} - 1}{N_{JK}}\right)  \sum^{N_{JK}}_{i = 1} (\xi_i(r) - \bar{\xi}(r))^2 },
 \end{equation} 
where $N_{JK}$ is the number of Jack-Knife subsamples used and $\xi_i(r)$ corresponds to the measurement of $\xi(r)$ excluding the $i$th subsample. 

The errorbars, computed with jacknife, give an idea of the scatter that we would expect from different realizations of the same volume and number density. However, in this study we compare different models run on the same haloes, and then these errors do not reflect the uncertainties of the scatter between the models. Differences between models are systematic and could be significant even when they are below the error bars. Then, differences below the errorbar must be taken with care in this study since the same comparison applied to a larger volume could reduce the errorbars but not necessarily the difference between the models.

\section{Results}\label{sec:results}

\begin{figure*}
\begin{centering}
\includegraphics[scale = .43]{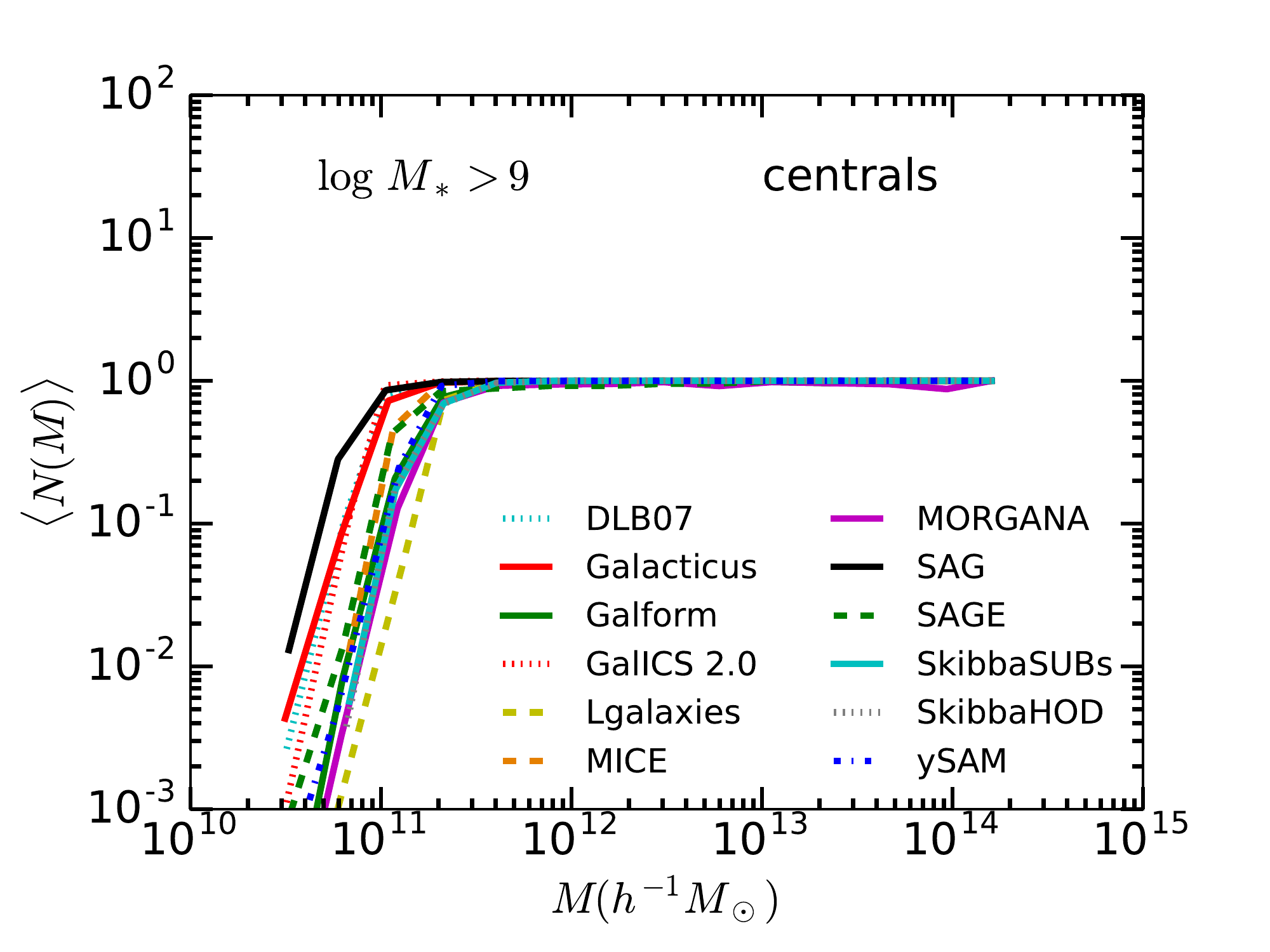}
\includegraphics[scale = .43]{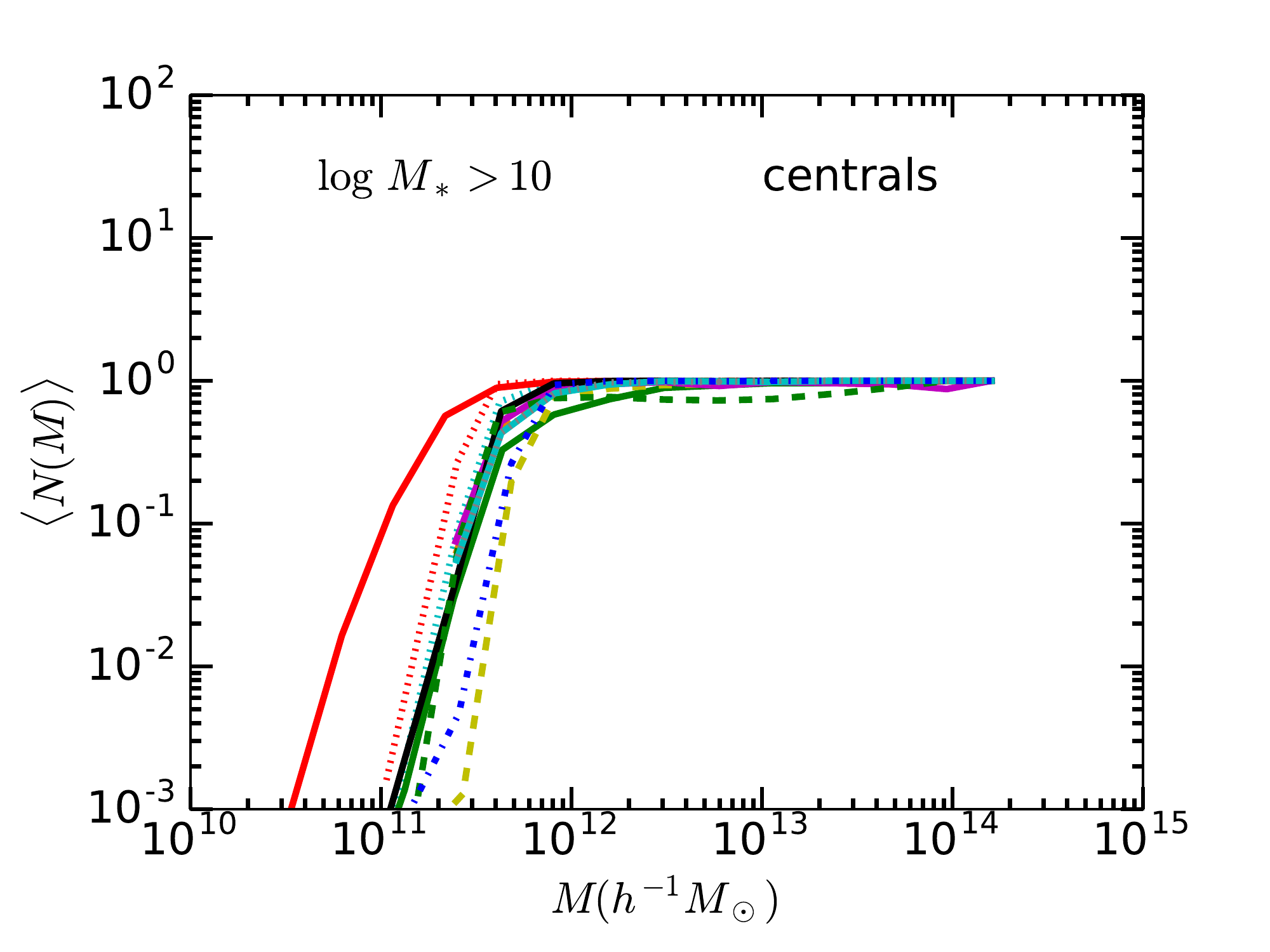}
\includegraphics[scale = .43]{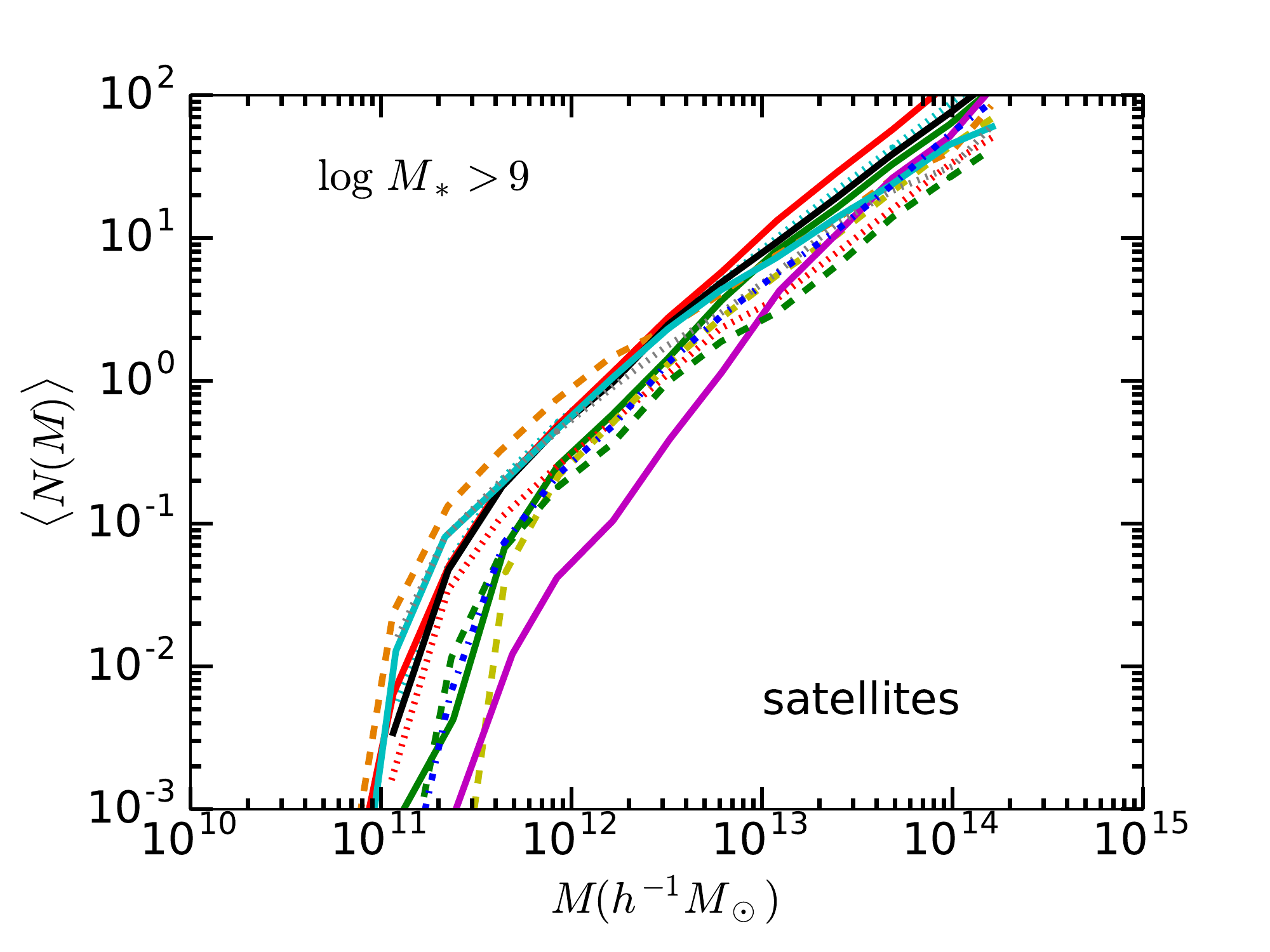}
\includegraphics[scale = .43]{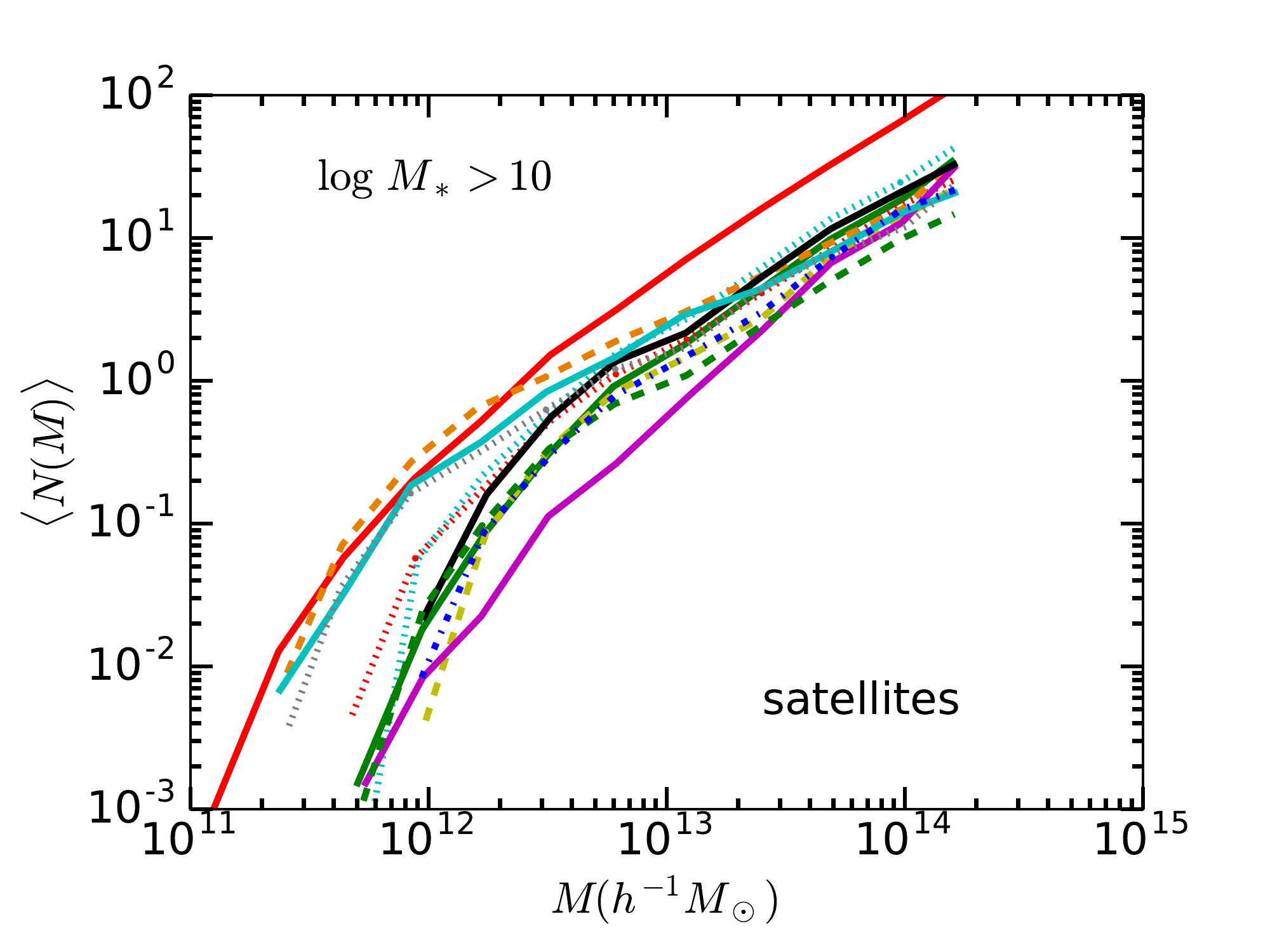}
\includegraphics[scale = .43]{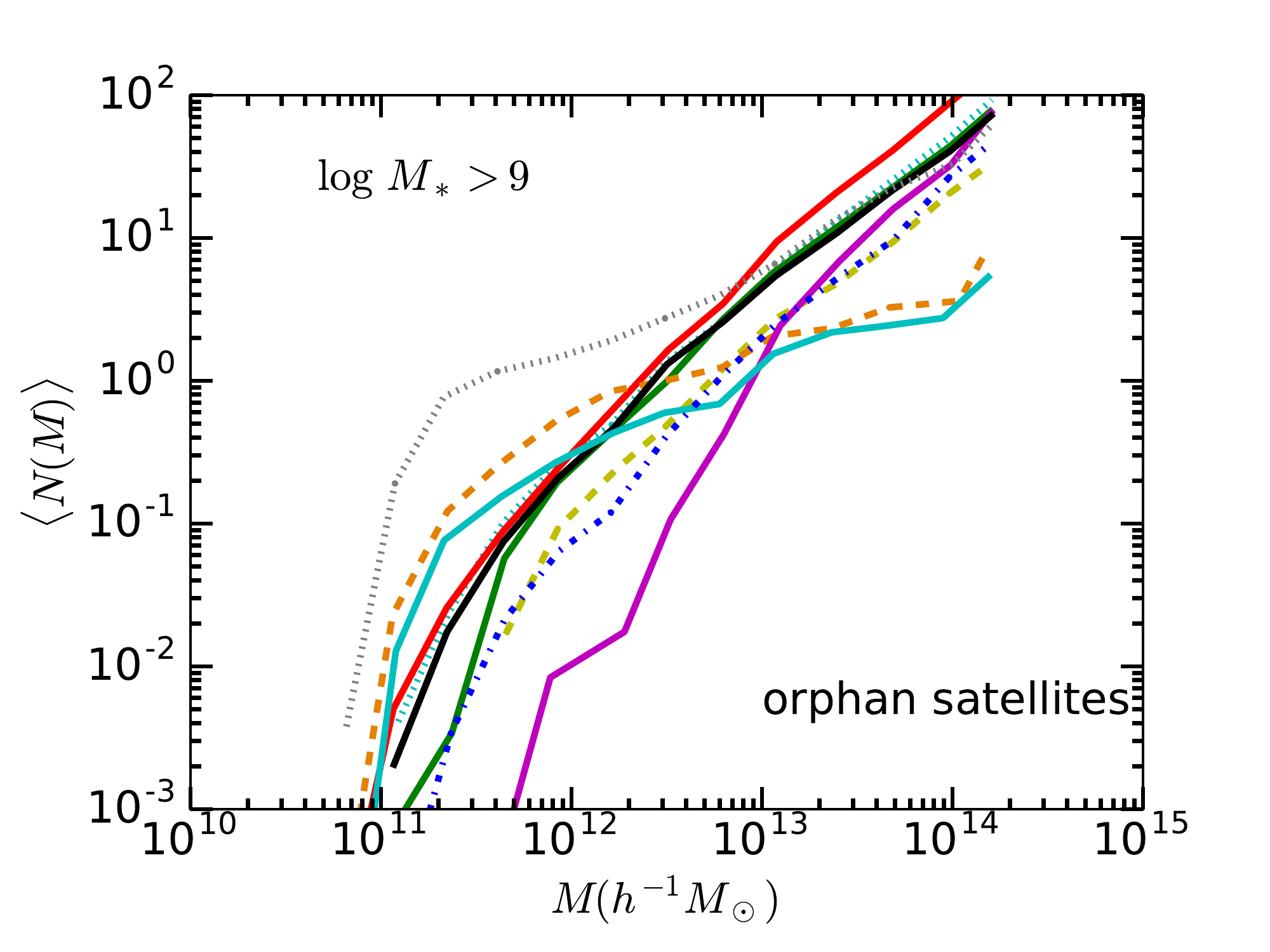}
\includegraphics[scale = .43]{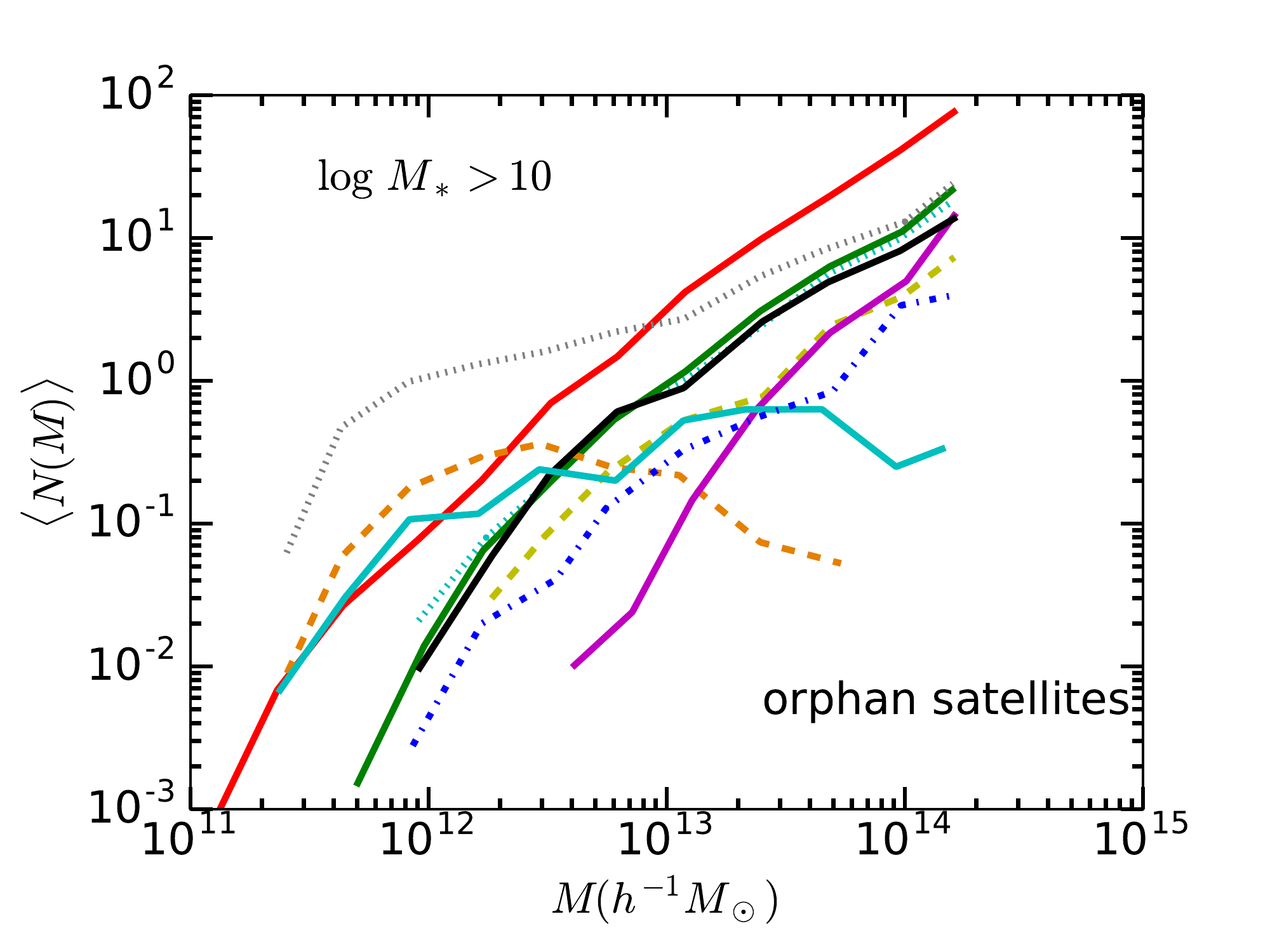}
\caption[Mean halo occupation number comparison of the different models]{Mean halo occupation number as a function of halo mass for the different models and galaxy types. Left panels show galaxies with $M_* > 10^9 \Mo$, while the right shows $M_* > 10^{10} \Mo$. Top panels show central galaxies, middle panels satellite galaxies and bottom panels orphan satellites. }
\label{fig:hod_comp_types}\par\end{centering}
\end{figure*}

In this section we present the model comparison of the  mean halo occupation number, the 2PCFs and the radial distribution of the galaxies in the haloes, with an emphasis on the orphan galaxy distribution. We use the galaxy catalogues at redshift $z=0$ and apply different stellar mass cuts in order to see the mass dependence of the convergence and differences of the models.



\subsection{Mean halo occupation number}\label{sec:hod}

In this section we study the mean number of galaxies populating haloes within a given range in mass. In the halo model paradigm it is usually assumed that haloes cluster according only to their masses. Hence, the distribution of galaxies provides a window into the clustering.

In both SAM and HOD models the galaxy populations are commonly characterized by central and satellite galaxies. According to these models, every halo can be occupied by at most one central galaxy, and only the haloes that contain a central galaxy can have a non-zero number of satellites:

\begin{equation}
 \langle N|M,M_\ast\rangle \,\equiv\, \langle N_\mathrm{cen}|M,M_\ast\rangle ( 1 +     \langle N_\mathrm{sat}|M,M_\ast\rangle ),
\end{equation}
with
\begin{equation}
\langle N_\mathrm{cen}|M,M_\ast\rangle <1,
\end{equation}
 where  $\langle N|M,M_\ast\rangle$ is the mean number of galaxies $N$ of stellar mass $M_*$ that populate haloes of mass $M$, and  $N_{cen}$ and $N_{sat}$ are the number of central and satellite galaxies respectively.
According to all this, for $\langle N | M \rangle \lesssim 1$ the contribution of the HOD mainly comes from the central galaxies and for $\langle N | M \rangle > 1$ the contribution mainly comes from the satellite galaxies.

In Fig. \ref{fig:hod_comp} we show the comparison of the mean halo occupation numbers of the models. This measurement corresponds to the mean number of galaxies per halo as a function of halo mass. Left panels show galaxies with $M_* > 10^9 \Mo$, and right panels show galaxies with $M_* > 10^{10} \Mo$. This figure gives similar information as Figures $13$ and $14$ from K15, where the number of galaxies is normalized by the halo mass to explore the specific frequency of galaxies as a function of halo mass.

We can see a scatter where each model starts populating galaxies for low stellar masses, which is a consequence of the different implementations of cooling, reionization and stellar feedback. In particular, the minimum mass where all the haloes are populated (i.e. where $\langle N(M) \rangle = 1$) changes a factor of $3$ between the models for galaxies of a stellar mass threshold of $M_* > 10^9 \Mo$. This scatter decreases to a factor of $2$ when the $M_* > 10^{10} \Mo$ cut is applied, except for \galacticus. 

If we focus on the right panel, we see a strong difference in \galacticus\ for $M_* > 10^{10} \Mo$, where these massive galaxies also populate very small haloes. This is due to the excess of galaxies at these masses for this model, that can be seen as a bump around $M_* \sim 2-3 \times 10^{10}\Mo$ in the stellar mass function from Fig. $2$ in K15. This comes from the fact that the galaxy formation model has been calibrated using another simulation to match observations. Changing the simulation without recalibrating the stellar mass function has a significant impact on \galacticus\ (see Fig. 6 of K15). We also like to mention that the fact that \skibbanfw\ and \skibba\ are not identical due to some stochastic components of the models. We also note that \sage\ and \galics\ show the lowest occupation number at high masses in the left panel. This is expected since these models do not have orphan satellites by construction.

In order to study the contributions of the different galaxy types, we show in Fig. \ref{fig:hod_comp_types}  the mean occupation number of galaxies split into galaxy types. Top panels show central galaxies, middle panels show satellite galaxies (orphan and non-orphan) and bottom panels show orphan satellites. The same stellar mass cuts as in Fig. \ref{fig:hod_comp} has been applied here for left and right panels.

Given the fact that all the haloes are populated by a central galaxy, the mean occupation number for a given stellar mass cut is directly related to the fraction of central galaxies that are more massive than the given stellar mass cut. Models whose central galaxies are less massive will present lower occupation numbers when a stellar mass cut is applied. 
If we focus on the top left panel where the central galaxies are shown for a stellar mass cut of $10^9\Mo$, we see that \sag, \galacticus, \galics\ and \dlb\ are the models that show higher occupation numbers, while \lgalaxy\ is the model that has the lowest values. This is consistent with Table $3$ of K15, where we see that \sag, \galacticus, \galics\ and \dlb\ are precisely the models that present more central galaxies above $10^9\Mo$, and \lgalaxy\ presents the lowest number.

If we focus on the middle panels, where satellite galaxies are plotted, we see that the models show a large scatter at small masses, but   the number of galaxies per halo increases with mass with a similar slope for larger masses. This agreement in the slope of the relation means that the galaxy formation models distribute satellite galaxies in haloes in a similar way. We note the different behaviour shown by \morgana, which shows a significantly lower occupation number for small haloes. This is due to the decoupled modelling of satellite galaxies with respect to substructures, than can leave some naked substructures without any satellite galaxy in it. The reason of this treatment of satellite galaxies is that the model was originally designed to be run in \textsc{Pinocchio} simulations, where only haloes (but not subhaloes) are obtained from the output of the simulations. 

\begin{figure*}
\begin{centering}
\includegraphics[scale = .43]{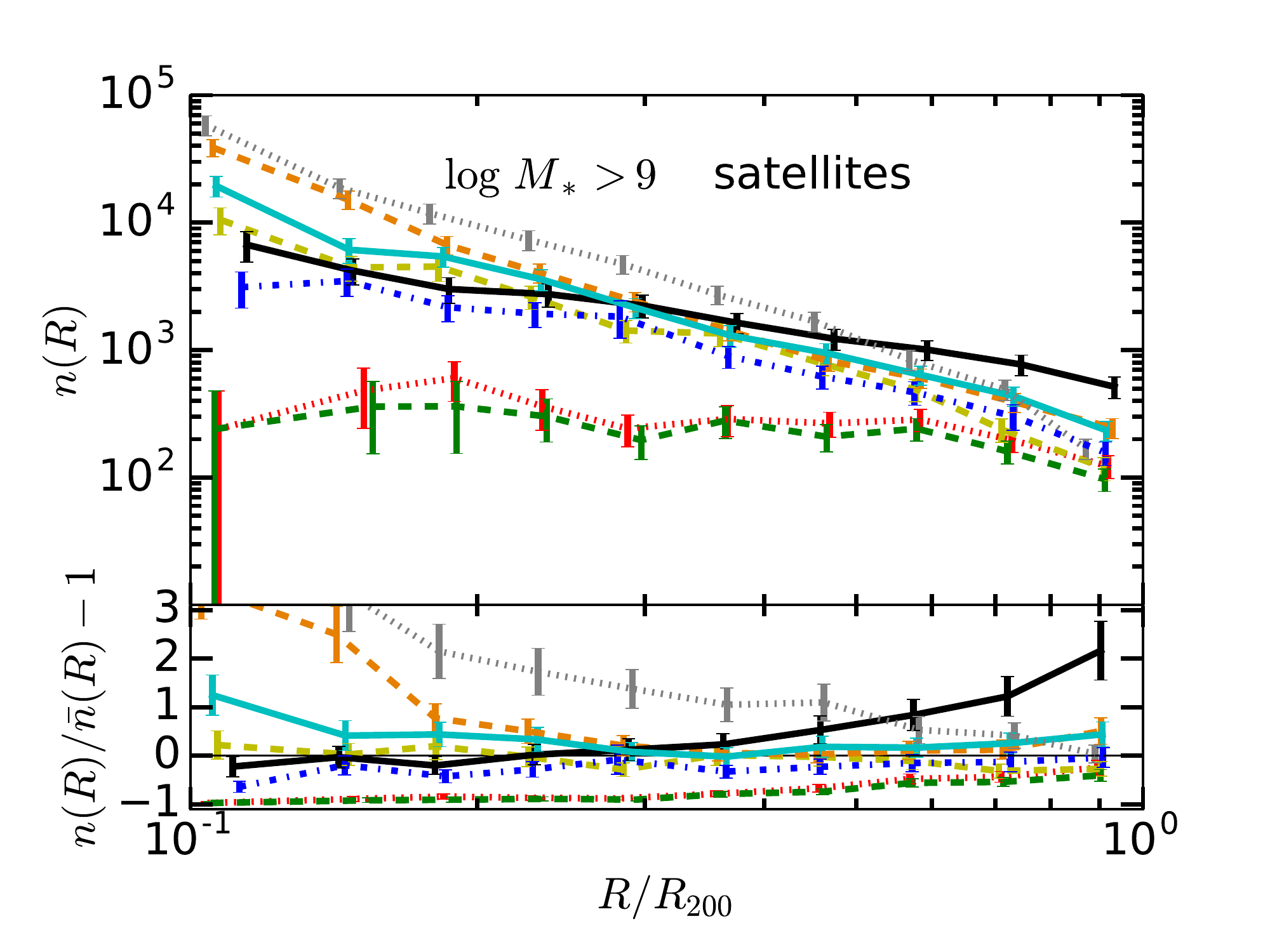}
\includegraphics[scale = .43]{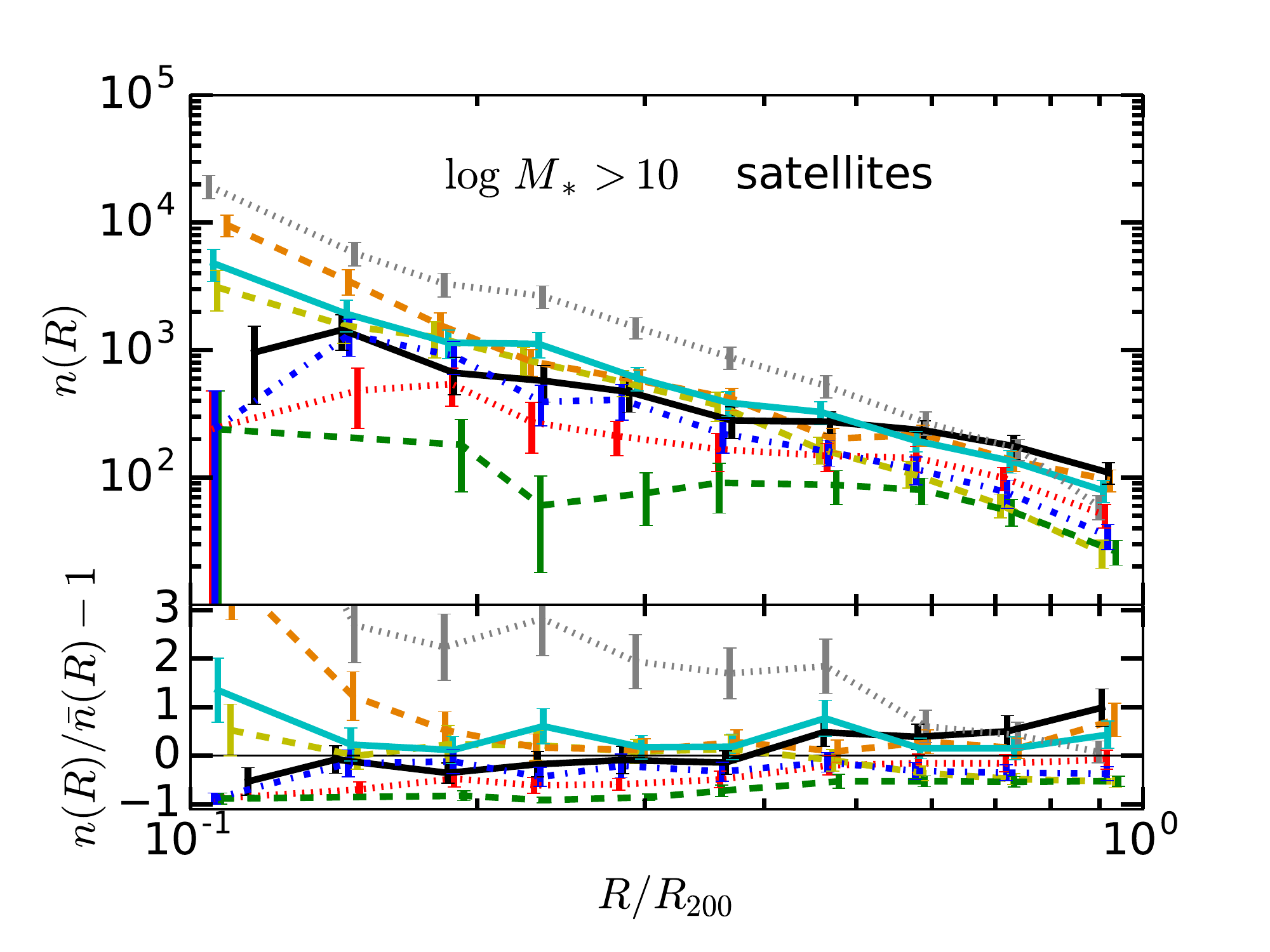}
\includegraphics[scale = .43]{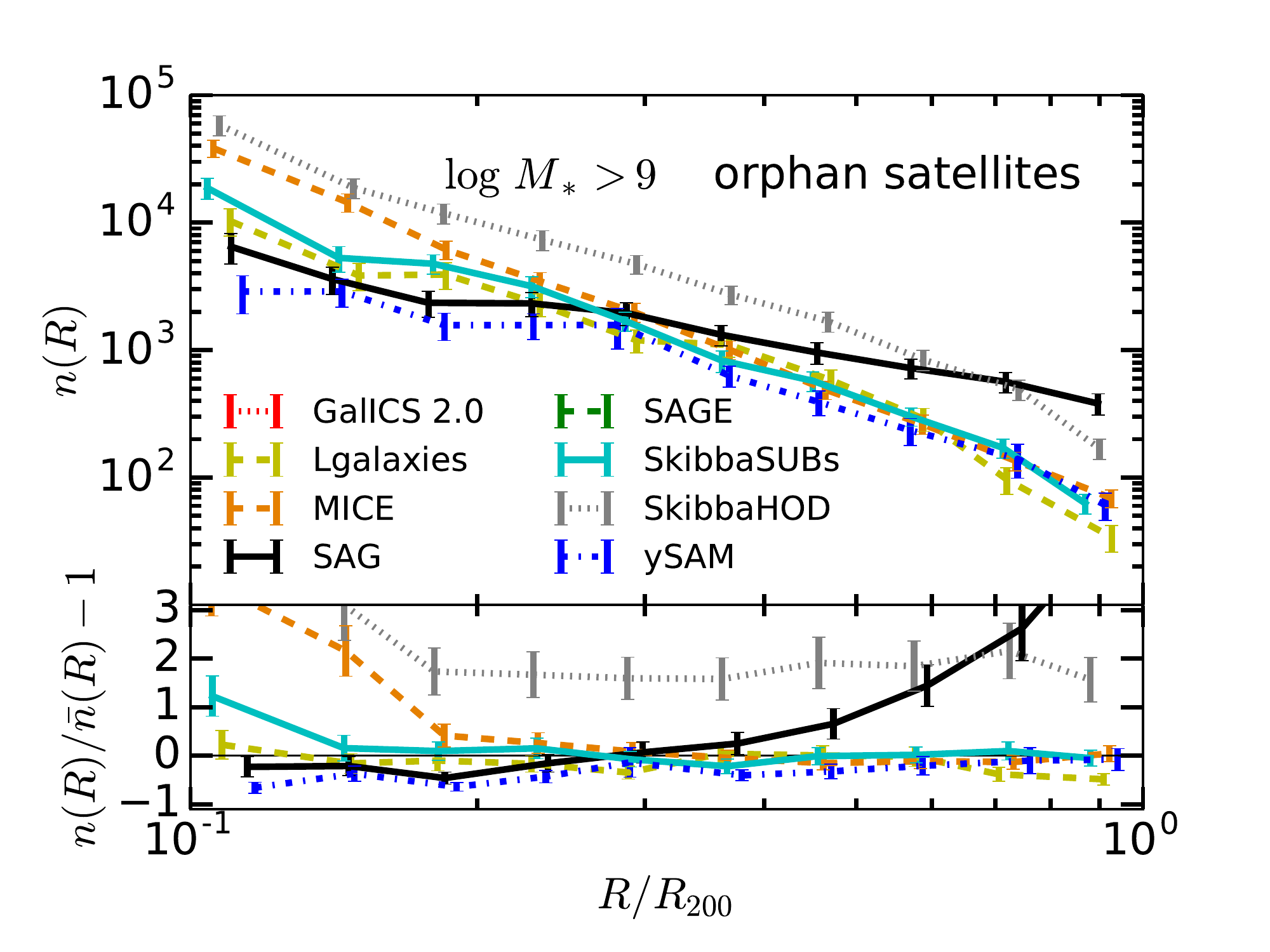}
\includegraphics[scale = .43]{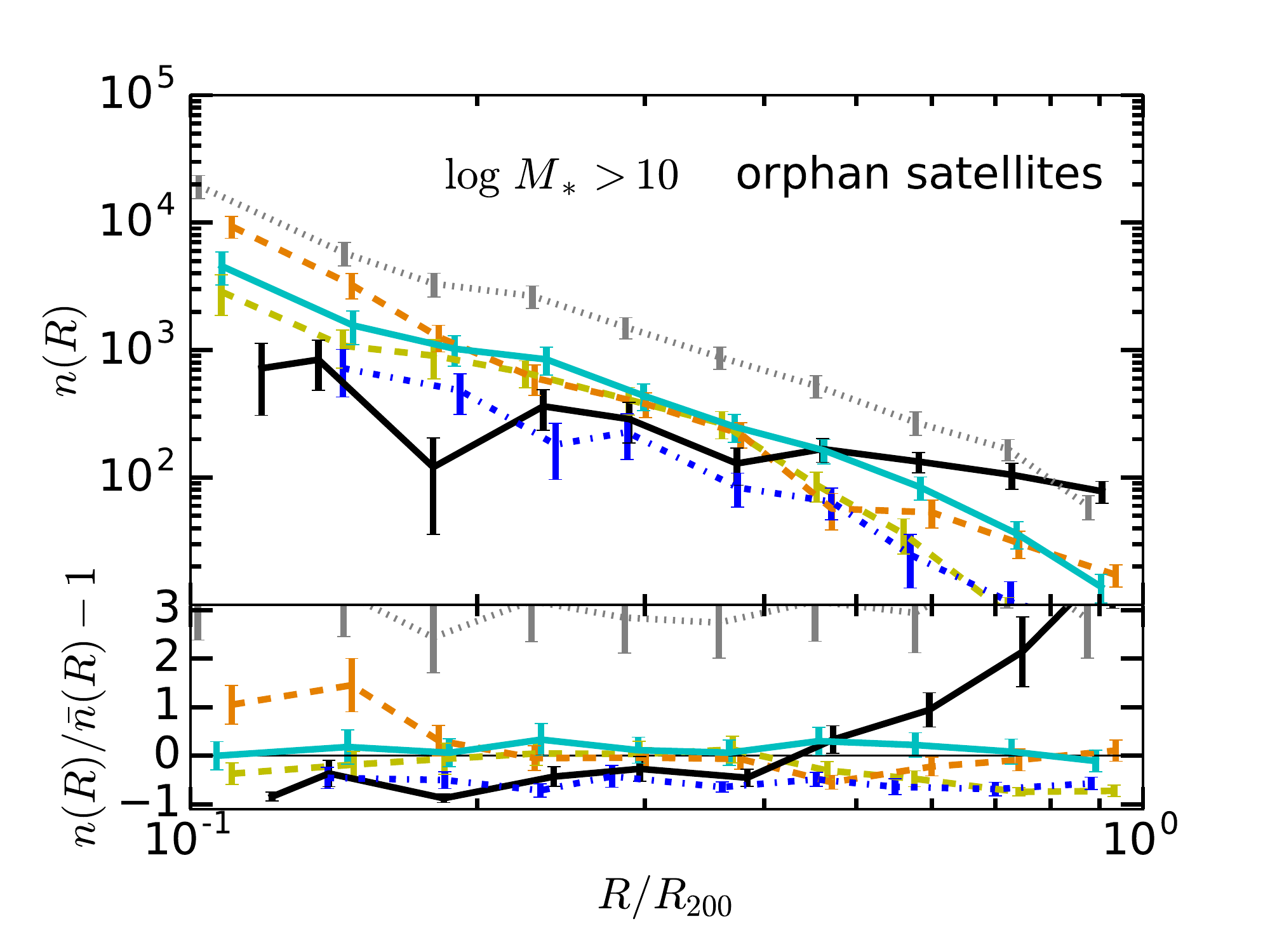}
\caption[Radial distribution comparison of the different models]{Comparison of the radial distributions of the different models. Left panels show galaxies with $M_* > 10^9 \Mo$, while in the right with $M_* > 10^{10} \Mo$. In the top panels we show all the satellite galaxies, while bottom panels show only orphan satellites. Each panel includes the residual with respect to the median of the distribution in each $R/R_{200}$ bin. }
\label{fig:dens_prof}\par\end{centering}
\end{figure*}


The differences between HOD models and SAMs are stronger on the bottom panels, where we show the mean halo occupation number for orphan satellites. We can see that the slope in the mean halo occupation numbers of HOD models is much shallower than SAMs for $M_* > 10^{12}\Mo$, showing a difference between HOD models and SAMs of one order of manitude higher at $M_* \approx 10^{14}\Mo$ than at $M_* \approx 10^{12}\Mo$. This is because massive haloes have many substructures and hence the HOD models occupy them with as many non-orphan satellites as possible. If the total occupation number is not high enough, then the number of orphan satellites in these haloes is low. On the contrary, orphan satellites in SAMs originate from the disruption of subhaloes, and this usually happens in high density environments with strong gravitational interactions. Consequently, in massive haloes, many subhaloes can interact with the environment and suffer tidal stripping. Hence, we expect that the orphan occupation increases quickly with halo mass, as we can see from these panels. 

Finally, note also that \skibbanfw\ has a higher amplitude than \skibba, and the slope is similar to some SAMs at high enough halo masses. This is due to the fact that \skibbanfw\ populates the haloes only with orphan satellites by construction, for which the total number is significantly higher than \skibba. Also see that most of the SAMs show a good agreement in this mean halo occupation number of orphan satellites for $M^* > 10^9\Mo$, specially for \dlb, \galacticus, \galform\ and \sag, and there is a good agreement between \lgalaxy\ and \ysam\ too. However, the differences become more significant for $M^* > 10^{10}\Mo$. \morgana\ shows the lowest occupation number due to the shorter merger times implemented in this model. 

\subsection{Radial distributions}\label{sec:dens_prof}

In this section we compare the radial distributions measured in all the models. We do the measurement from the following equation:
\begin{equation}
n(R/R_{200}) = \frac{N_g}{(4 \pi /3) [(R + \Delta R)^3 - R^3]},
\end{equation}
where $n(R/R_{200})$ is the number density of galaxies in the radial annulus $R$ to $R + \Delta R$, with $R$ referring to the radial distance to the halo centre, and $N_g$ is the number of galaxies between $R$ and $R + \Delta R$. $R_{200}$ is the radius that encloses $200$ times the critical density. So, this basically describes the density of galaxies as a function of the radial distance of the halo centre. 

Fig. \ref{fig:dens_prof} shows the radial distribution of the different models, applying the same stellar mass thresholds of $M_* > 10^9 \Mo$ (left) and $M_* > 10^{10} \Mo$ (right) as in the previous figure. Top panels show all the satellites, while bottom panels show only orphan satellites. Each panel shows the residual as $n(R)/\bar{n}(R) - 1$, where $\bar{n}(R)$ is the median of the distribution at each $R$ bin. 

\begin{figure*}
\begin{centering}
\includegraphics[scale = .43]{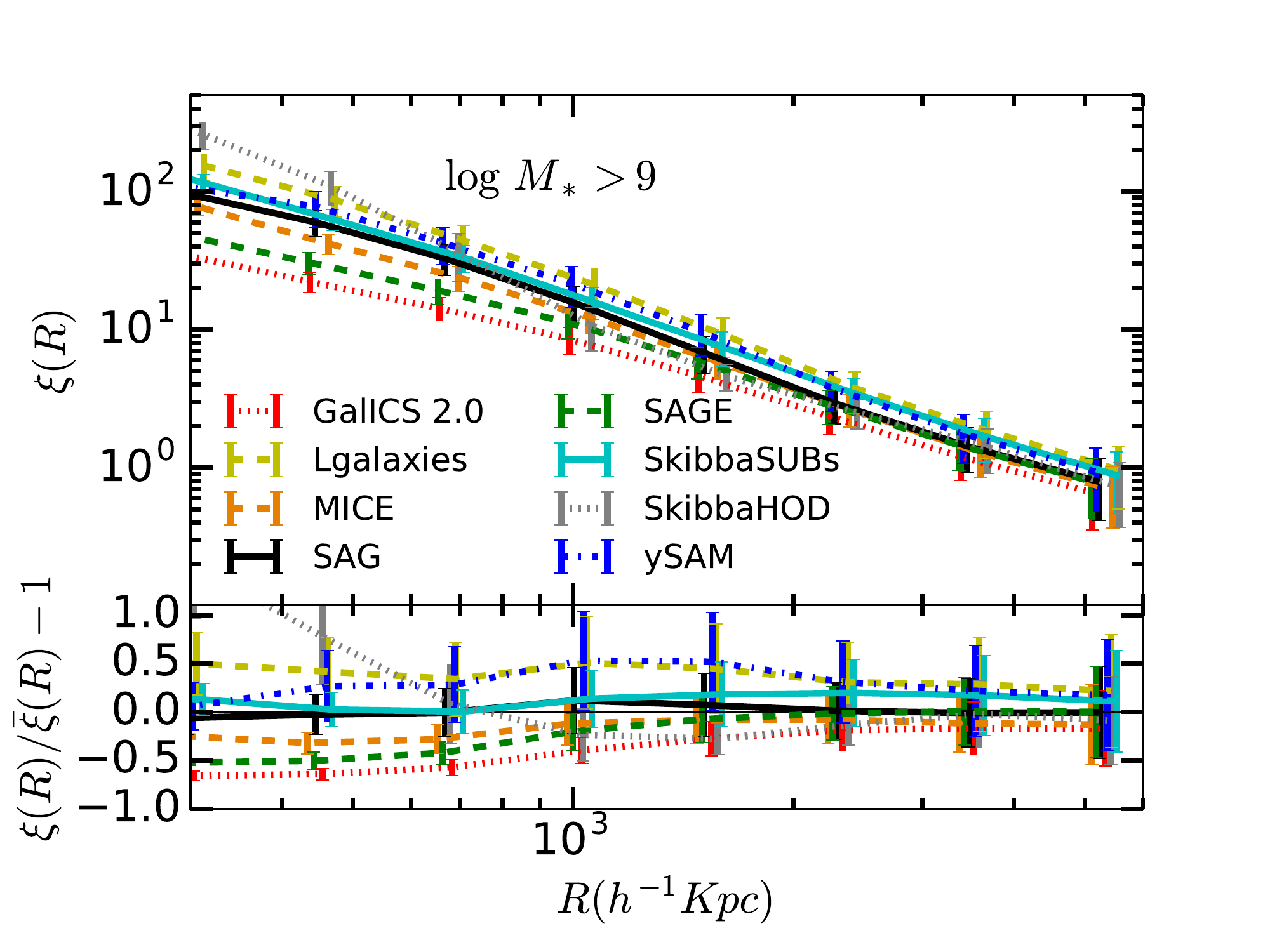}
\includegraphics[scale = .43]{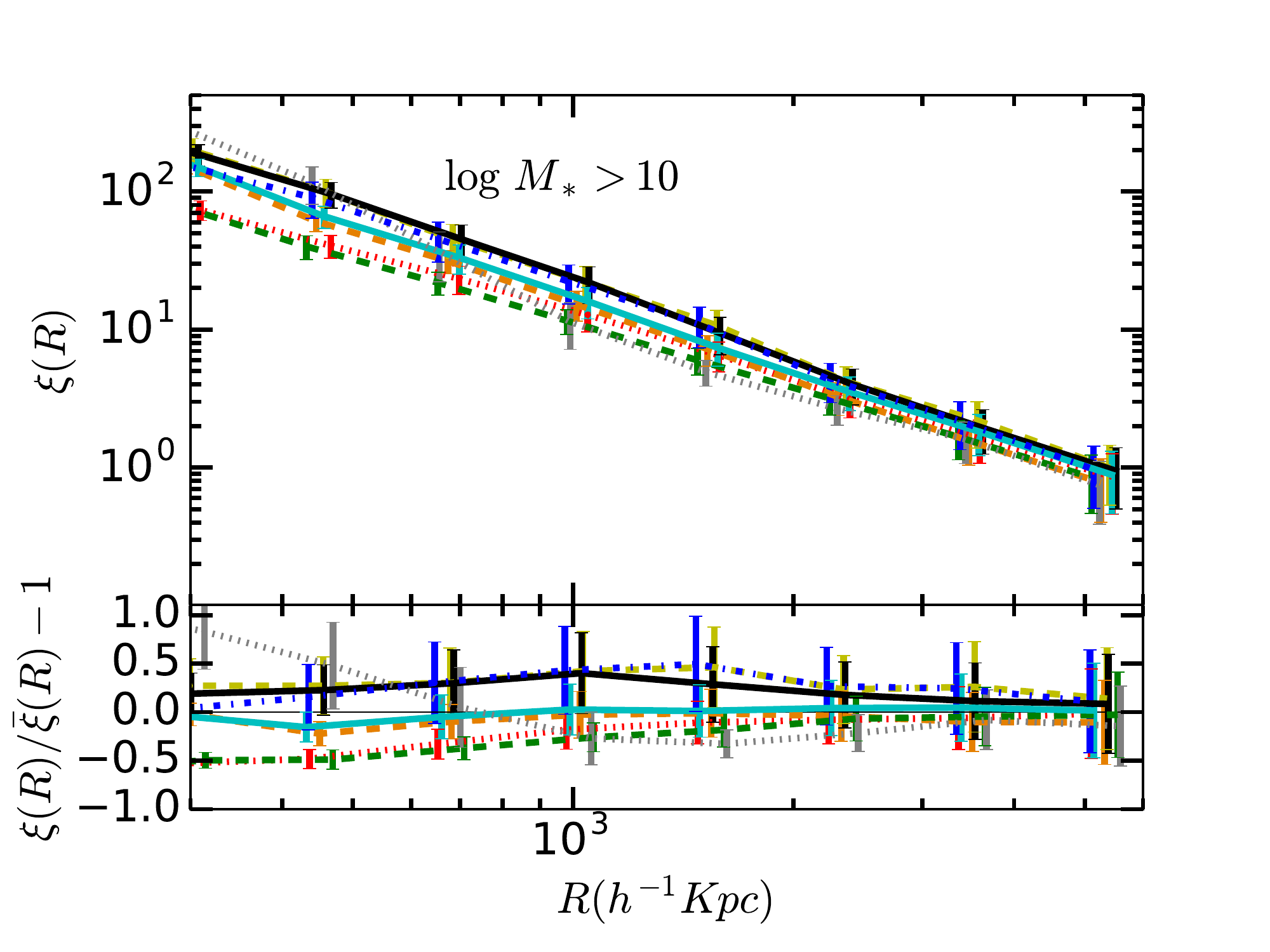}
\caption[2PCF comparison of the different models]{Comparison of the 2PCF measurements of the different galaxy formation models using a stellar mass threshold. Left panels show galaxies with $M_* > 10^9 \Mo$, while right panels correspond to $M_* > 10^{10} \Mo$. Both panels use all the galaxies of the models. }
\label{fig:2pcf_mstar_all}
\par\end{centering}
\end{figure*}

\begin{figure*}
\begin{centering}
\includegraphics[scale = .43]{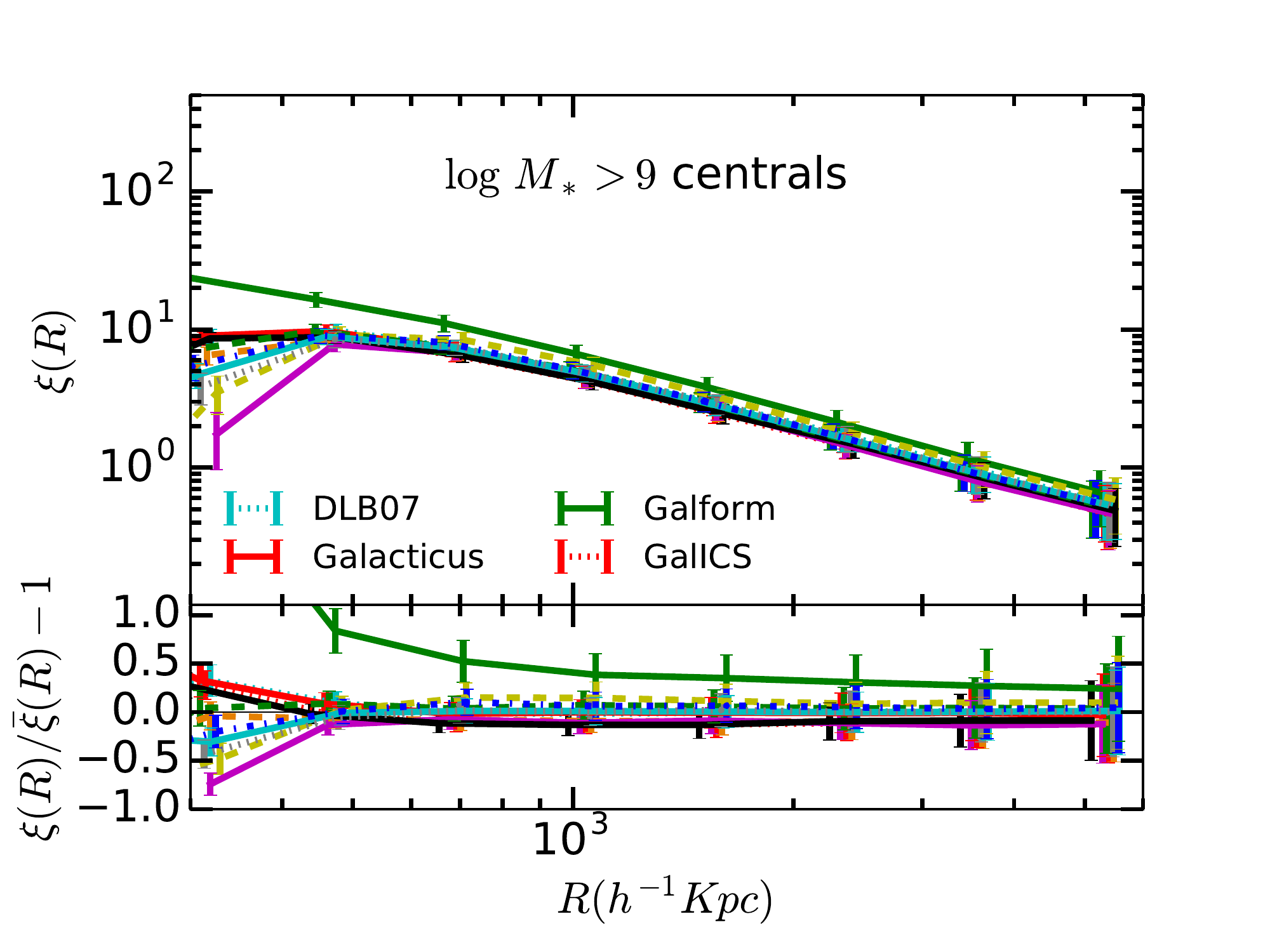}
\includegraphics[scale = .43]{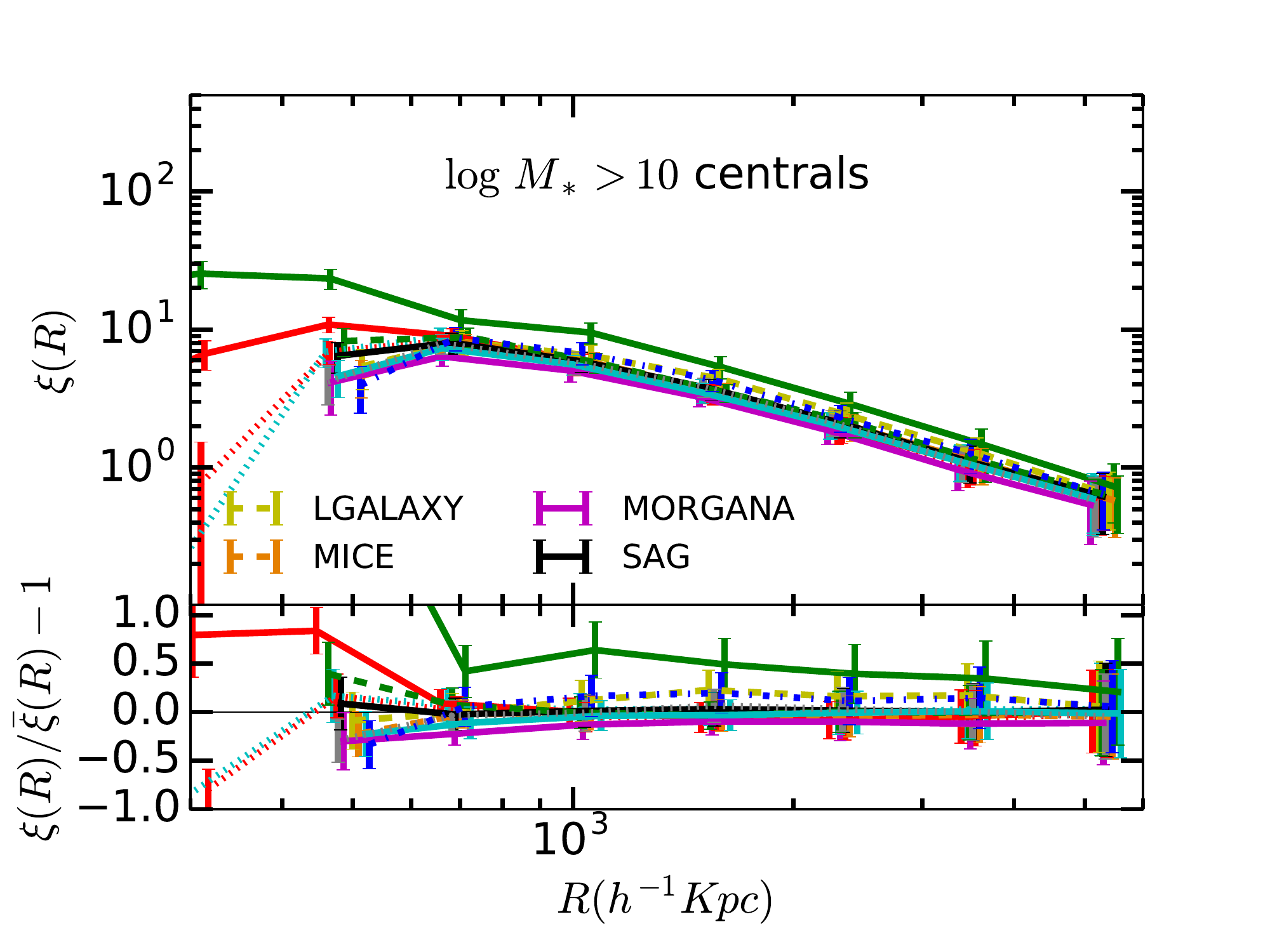}
\includegraphics[scale = .43]{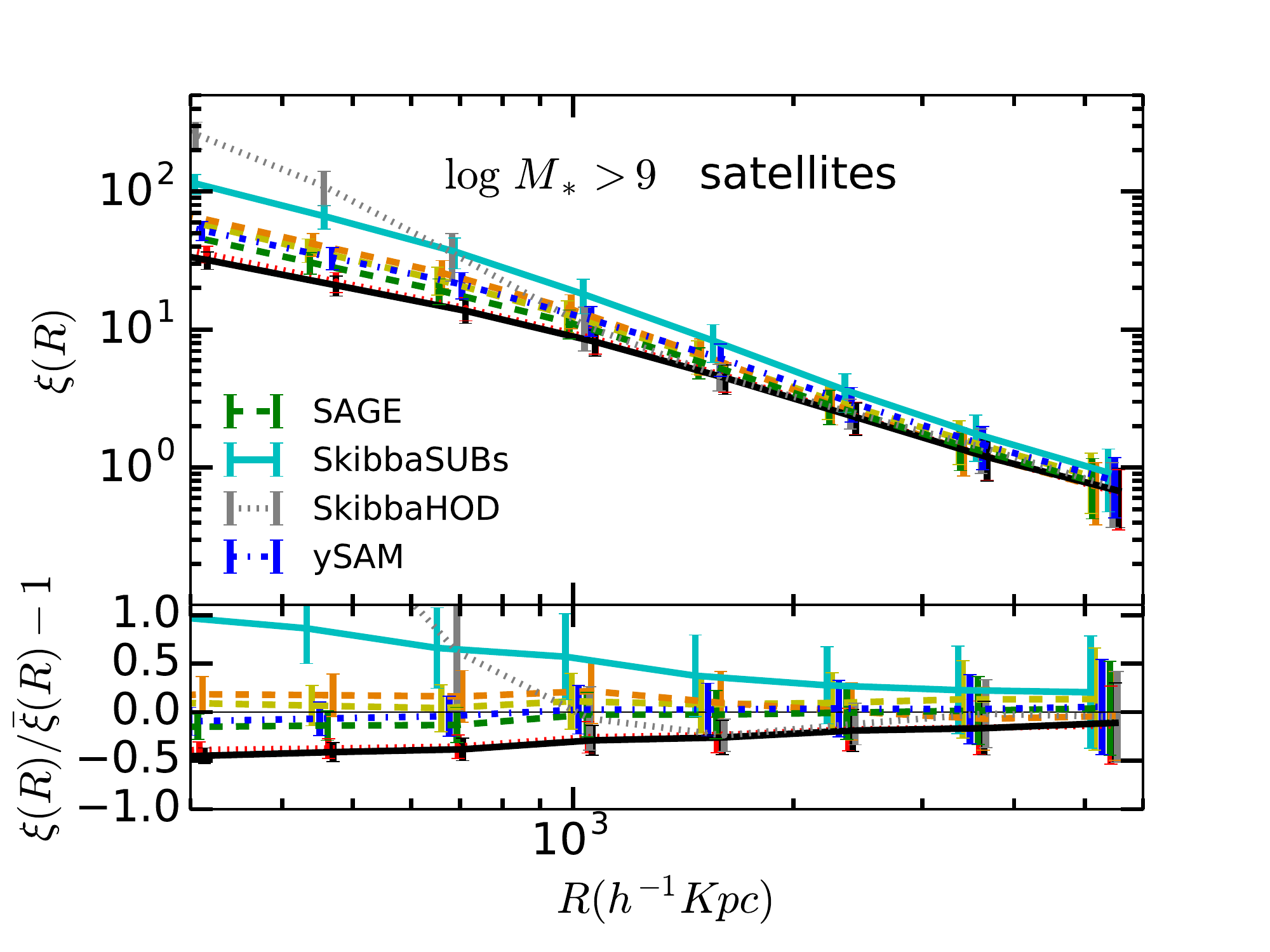}
\includegraphics[scale = .43]{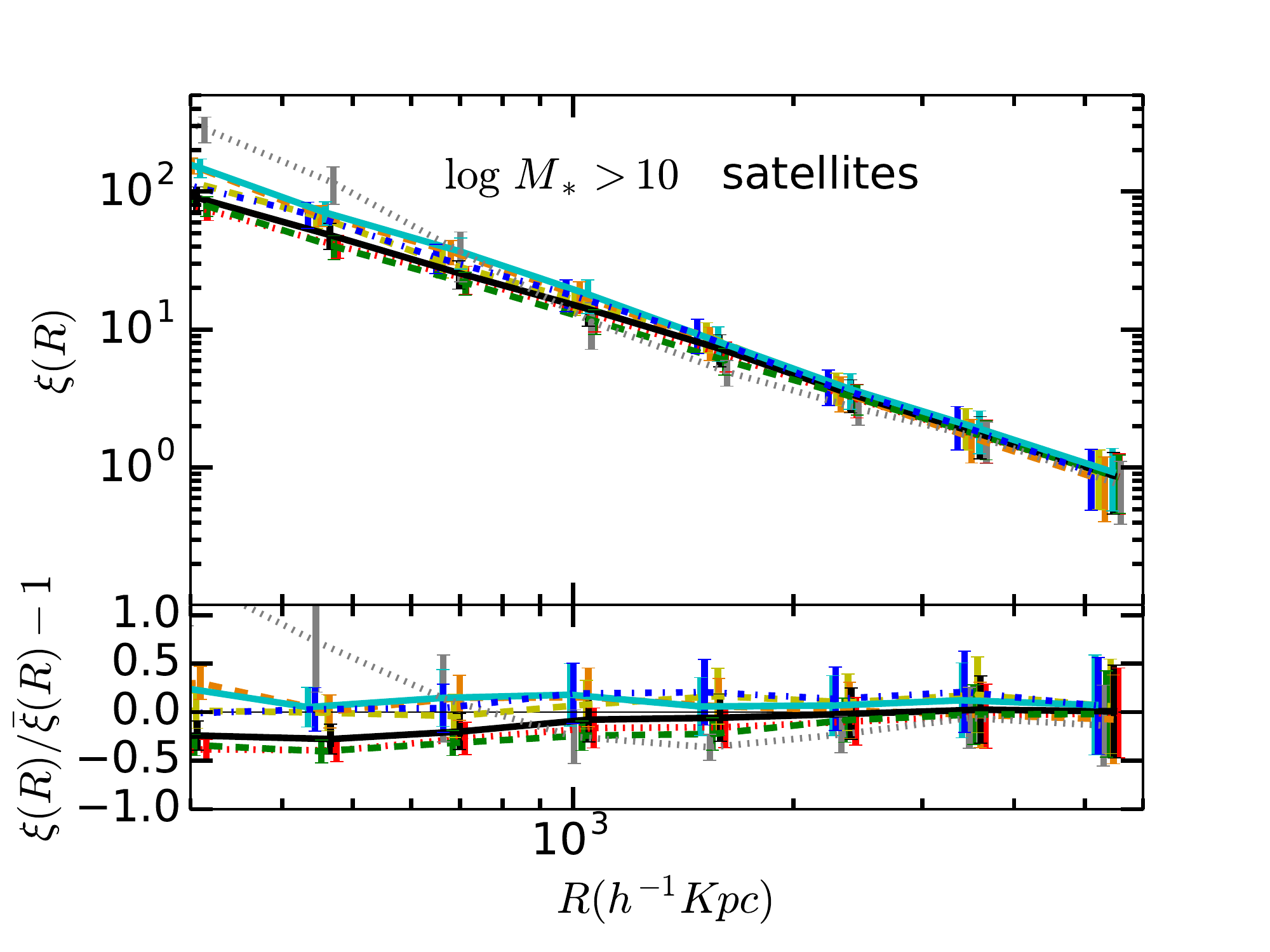}
\includegraphics[scale = .43]{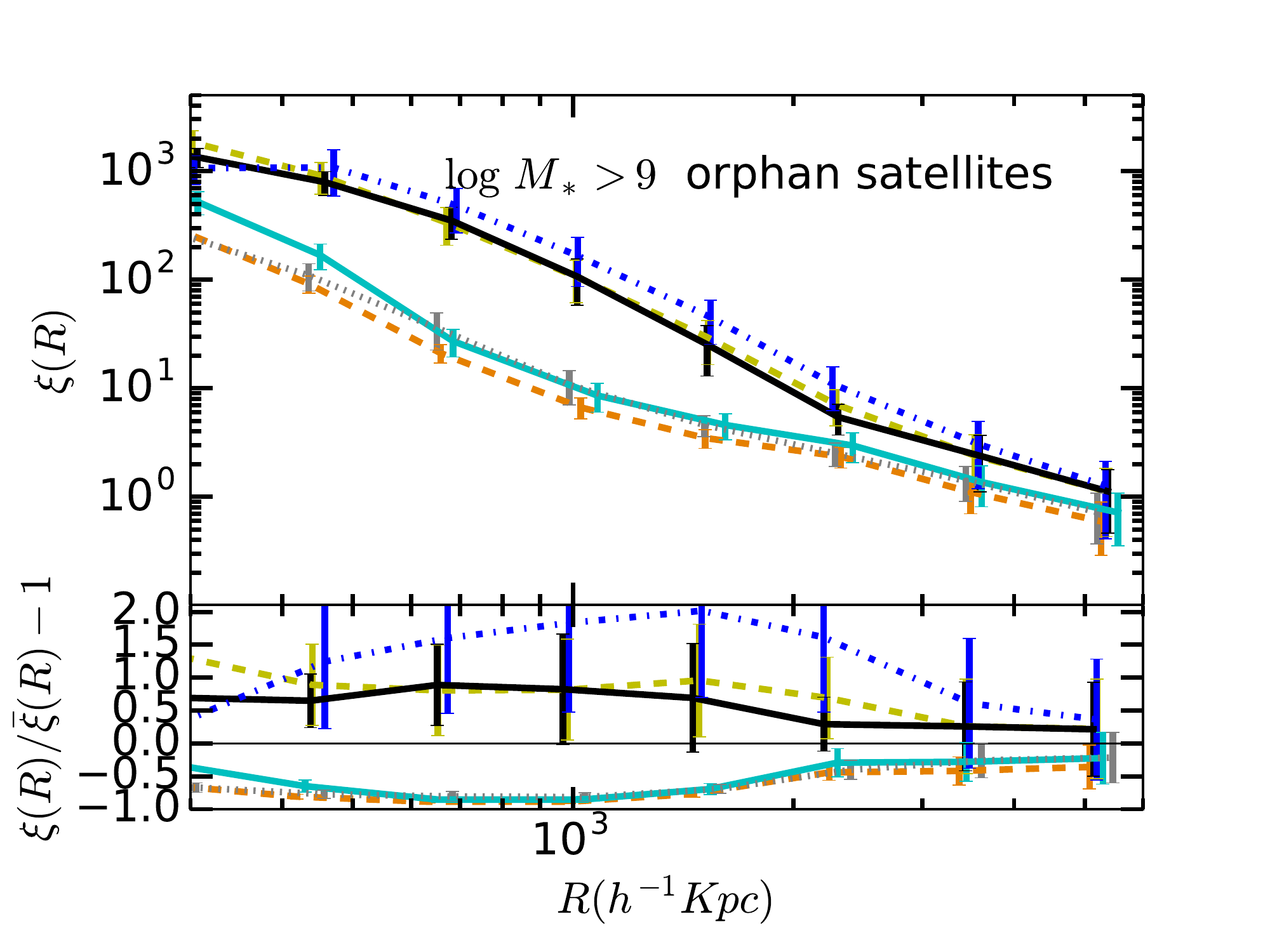}
\includegraphics[scale = .43]{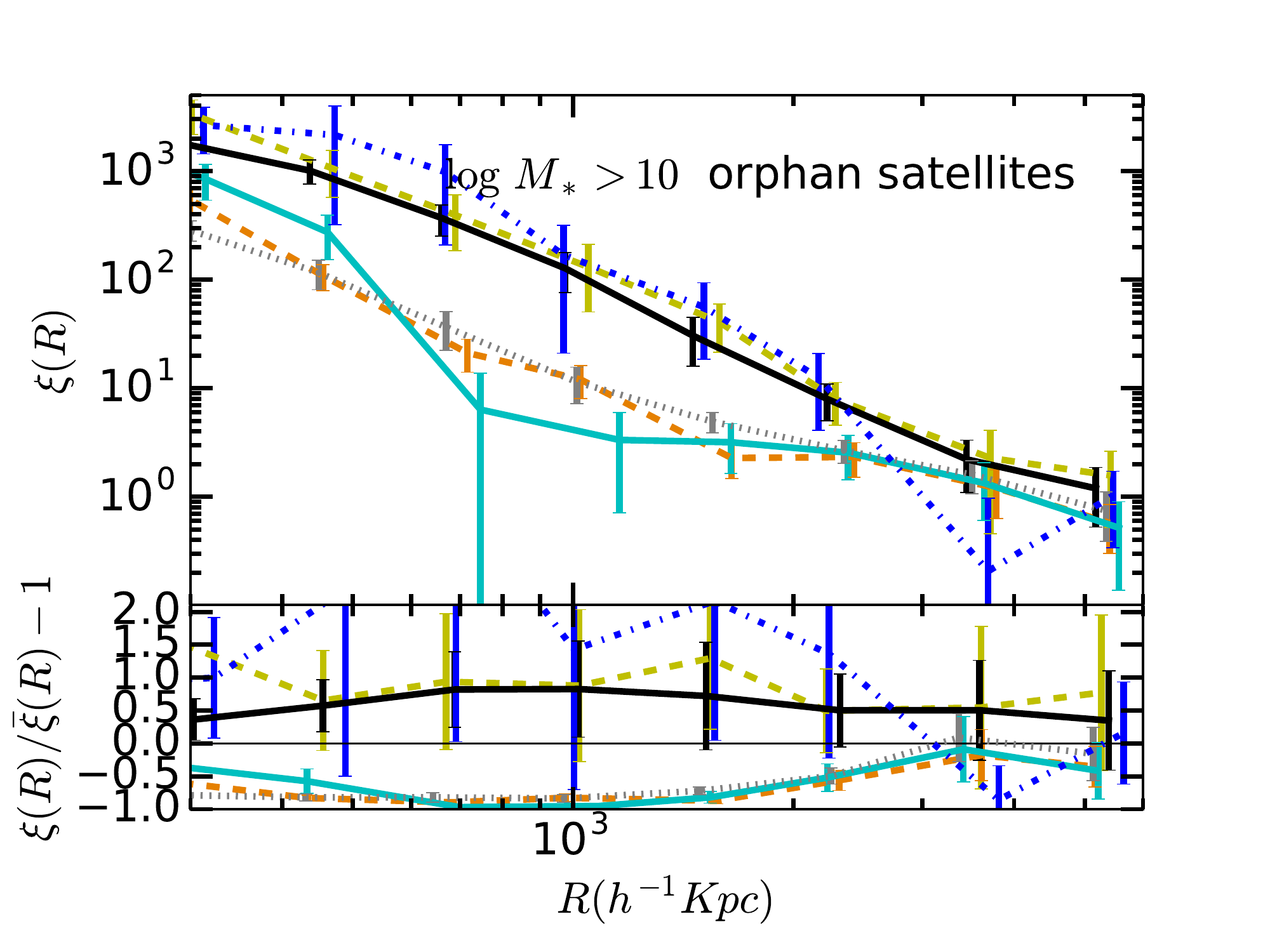}
\caption[2PCF comparison of the different models]{Comparison of the 2PCF measurements of the different galaxy formation models using a stellar mass threshold. Left panels show galaxies with $M_* > 10^9 \Mo$, while in the right shows galaxies with $M_* > 10^{10} \Mo$. In the top panels we show the central galaxies, middle panels all the satellite galaxies, and bottom panels orphan satellites. }
\label{fig:2pcf_mstar_types}
\par\end{centering}
\end{figure*}

In the top panels we only show those models that have implemented a treatment for orphan satellites (or have not done it by construction). The differences between most of the models are lower than $2\sigma$ for $M_* > 10^9 \Mo$ and lower than $1\sigma$ for $M_* > 10^{10} \Mo$. We see that the scatter is smaller than a factor of $2$ for all scales smaller than $0.2 R/R_{200}$ and for all the models except \skibbanfw, \galics\ and \sage. \skibbanfw\ shows the highest radial distribution, a factor of $2$ higher than the median. This is because all the satellites are orphans, and hence all the galaxies follow a NFW profile instead of following substructures. The fact that most of the models agree with the radial distribution of \skibbanfw\ except for the smallest scales is consistent with previous studies \citep{Gao2004,Kang2014,Pujol2014b,vanDaalen2016}.
On the other hand, \sage\ and \galics\ show much flatter radial distributions than the rest of the models. This is because these models have no orphans, and hence only non-orphan galaxies contribute to these distributions. These two models then show the contribution of non-orphan satellite galaxies to the radial distributions, showing agreement with the rest of the models only at the largest scales. 

We see that the HOD models (\mice, \skibbanfw\ and \skibba) present a steeper slope of the radial distributions than SAMs in all the cases. Given the similarities between the top and bottom panels, we see that these differences basically come from the different treatments of orphan satellites, which dominate the smallest scales (satellites only contribute to the large scales of the panels). This indicates that the orphan satellites have a very important role in the distribution of galaxies in haloes, consistent with the conclusions at  \cite{Gao2004}.  While HOD models distribute orphan satellites without any information from substructure or evolution (in \skibbanfw\ this is the case for all satellites), orphan satellites from SAMs are a consequence of subhalo disruption, and hence the positions of orphan satellites are correlated with substructure. 
Moreover, orphan satellites in SAMs are limited in the densest regions, close to the halo centre, where orphans merge quickly with the central galaxy. This exclusion effect in the inner parts of the halo is one reason of why SAMs show a lower orphan density at the smallest scales.

\subsection{Two point galaxy correlation functions}\label{sec:2pt}

We now compare the 2PCF between the different models. We have applied the same stellar mass thresholds used previously, and we also study the different galaxy types separately. Again, we show the residuals with respect to the median of the measurements in order to see the scatter between the models. Since the stellar mass functions are different between the models, the number density of galaxies for the same stellar mass cut can be different. We have also studied the differences using number density cuts instead of stellar mass cuts and we obtain the same results, meaning that the difference between the models is not due to the differences between their number densities.

In Fig. \ref{fig:2pcf_mstar_all} we show the 2PCFs of galaxies for all the models that have computed the orphan positions (and those which did not do it by construction), using all the galaxies.  For both stellar mass cuts we find a good agreement, with most of the models consistent within the error bars (although we must be carefull when interpreting the errorbars, as discussed in \S\ref{sec:methodology}). This is an encouraging result, since it highlights a consistency between the models even when most of them did not use any observations of clustering to constrain their parameters. However, we note that \galics\ and \sage\ show a lower clustering at small scales, showing a factor of $2$ lower at the smallest scales. Again, this is a consequence of the fact that these models do not have orphan satellites. These models then show the impact that excluding orphan satellites can have on the galaxy clustering predictions.

We split the galaxy samples according to their type in Fig. \ref{fig:2pcf_mstar_types}.  We show in the top and middle panels the 2PCF of galaxies for central and satellite galaxies respectively.  In the cases of central galaxies the models have very good agreement, while some differences appear for satellite galaxies. The scatter between models in the middle panels is larger for smaller stellar masses, since galaxies are more dominated by orphan galaxies. In particular, all the satellites in \skibbanfw\ are orphans, causing a large difference with the rest of the models at the smallest scales due to the orphan radial distributions discussed previously. We also note a high clustering signal for \skibba, and a lower signal for \galics\ (which has no orphans) and \sag. For most of the models, the scatter between the models is in general lower than a $25\%$ for the satellite galaxies and lower than $20\%$ for the central galaxies. 

In the bottom panels of Fig. \ref{fig:2pcf_mstar_types} we focus on the models that have orphan satellites in order to study their distribution. For both stellar mass thresholds we see a strong and significant difference between the HOD models and SAMs. The HOD models show a lower clustering amplitude, and they all agree between them, while SAMs agree between them but with a higher amplitude and different shape than the HODs. 
The clustering of 2PCF on small scales depends on two main factors, the halo occupation number and the density profile of galaxies in haloes, especially in massive haloes. Although the HOD models show a slightly steeper radial distribution of orphan satellites with respect to SAMs, they also show a much flatter halo occupation number distribution, as indicated in the lower panel of Fig. \ref{fig:hod_comp_types}. This implies that orphan satellites in HODs populate less massive haloes than in SAMs. This is the main reason for the lower 2PCF of orphan satellites in HOD models compared to SAMs, since small scale clustering is strongly affected by the occupation numbers in massive haloes. 
This difference between the clustering of orphan satellites of SAMs and HOD models is large, reaching an order of magnitude at scales of $\approx 1 \Mpc$. 
The impact of orphan satellites in the agreement between models will depend in general on the orphan fraction of the galaxy samples, and we have seen that it also depends on the galaxy formation model. In this analysis the models that computed the orbits and positions of the galaxies are those with the lowest orphan fractions, meaning that the impact of orphan satellites on the clustering of the rest of the models might be stronger.

\section{Discussion and conclusions}\label{sec:conclusions}

In this article, we present a comparison of the clustering and halo occupation statistics of 12 different galaxy formation models. We use a dark matter only N-body simulation and run SAMs and HOD-based models with the same dark matter and merger trees input, and compare the results of mean halo occupation numbers, radial distributions of galaxies in haloes and 2-Point Correlation Functions (2PCFs). The goal of this paper is to study the clustering and distribution of galaxies in haloes, and to understand the roles of different galaxy types, in particular of orphan satellites (satellites which are not assigned to any dark matter subhalo). This work is part of a series of papers comparing galaxy formation models that started with K15.


The most important results of the study can be summarized as follows:


(1) The slope in the mean occupation number of orphan satellites as a function of halo mass is much shallower in HOD models than in SAMs, due to the different treatments of orphan satellites between both approaches. Orphan satellites in SAMs originate from the disruption of subhaloes, and this happens more often in massive haloes. However, in this study most of the HOD models populate satellites in subhaloes and only when there are more satellites than subhaloes these extra galaxies are considered as orphans. As massive haloes have many subhaloes, the number of orphan satellites in these HOD models is not as high as in SAMs. 


(2) HOD models have a steeper radial distribution of orphan satellites in haloes than SAMs. This is because HOD models distribute orphan satellites following a NFW profile, independently of the substructure and evolution of the haloes. This allows HOD models to populate with more orphan satellites in the inner and denser regions of the haloes than SAMs. SAMs are constrained to where subhaloes have been disrupted, and this causes a lower density in the innermost regions, where subhaloes quickly merge into the central structure, as well as galaxies  merging with other galaxies for some models. When comparing all the satellites, the different models have a scatter of $2$ times the measurement uncertainty (due to the limited volume used) in their radial distribution of galaxies for $M_* > 10^9 \Mo$. The scatter is smaller for higher thresholds, since orphan satellites become less relevant. 

(3) Using all the galaxies above a certain mass threshold for the measurements of 2PCFs (see Fig. \ref{fig:2pcf_mstar_all}), we see a scatter of a factor of $2$ between the models. However, part of this scatter is due to the lower clustering found for the models that do not have orphan galaxies by construction, an indication of the importance of orphan galaxies on galaxy clustering. We find a good agreement between the models for central galaxies and at large scales for all the galaxy selections. Using a larger volume in simulations would allow us to measure linear bias, and this would be a valuable extension to this work.

(4) HOD models and SAMs have significant differences in their clustering of orphan satellites (see bottom panels of Fig. \ref{fig:2pcf_mstar_types}). Both SAMs and HODs show good agreement for models of the same kind, but SAMs have a higher 2PCF than HOD models. This is due to the differences on the halo occupation numbers of orphan satellites between both schemes. Although HOD models show a steeper radial distribution for orphan galaxies than SAMs, they statistically occupy less massive haloes. The clustering at small scales is strongly affected by the halo occupation of massive haloes, and because of this the orphan satellites in SAMs show a  higher 2PCF at these scales than HOD models. 

It is important to notice that the models used have not been re-calibrated for this particular simulation. The agreement between the models could be improved by calibrating the models in the simulation where the comparison has been done or even using the same observational constraints (Knebe et al., in prep.). 
This study is limited by the resolution of the simulation. A higher resolution simulation would allow us to study smaller scales, and would also allow to detect subhaloes in inner regions of the haloes. This could have an impact on both the satellite distributions at small scales. In addition, a comparison with hydro-dynamic simulations would be useful to study the baryonic effects on both galaxy and dark matter clustering. It has been shown that baryons affect the dark matter distribution at small scales \citep{Tissera2010,Sawala2013,Cui2012,Cui2014,Cui2016}. Finally, another interesting extension would be the comparison of SAMs with new implementations that take into account observations of galaxy clustering to constrain their parameters \citep{vanDaalen2016}.

\section*{Acknowledgements}

The authors would like to express special thanks to the Instituto de F\'isica Te\'orica (IFT-UAM/CSIC in Madrid) for its hospitality and support, via the Centro de Excelencia Severo Ochoa Program under Grant no. SEV-2012-0249, during the three week workshop 'nIFTy Cosmology' where this work developed. We further acknowledge the financial support of the 2014 University of Western Australia Research Collaboration Award for 'Fast Approximate Synthetic Universes for the SKA', the ARC Centre of Excellence for All Sky Astrophysics (CAASTRO) grant number CE110001020, and the two ARC Discovery Projects DP130100117 and DP140100198. We also recognize support from the Universidad Autonoma de Madrid (UAM) for the workshop infrastructure.

Funding for this project was partially provided by the Spanish Ministerio de Ciencia e Innovaci\'{o}n (MICINN), Consolider-Ingenio CSD2007- 00060, European Commission Marie Curie Initial Training Network CosmoComp (PITNGA-2009-238356). We acknowledge support from the European Commission's Framework Programme 7, through the Marie Curie International Research Staff Exchange Scheme LACEGAL (PIRSES-GA-2010-269264).
AP acknowledges support from beca FI and 2009-SGR-1398 from Generalitat de Catalunya, project AYA2012-39620 and AYA2015-71825 from MICINN, and from a European Research Council Starting Grant (LENA-678282). 
RAS acknowledges support from the NSF grant AST-1055081. 
FJC acknowledges support from the Spanish Ministerio de Econom\'{i}a y Competitividad project AYA2012-39620. 
SAC acknowledges grants from CONICET (PIP-220), Argentina. 
DJC acknowledges receipt of a QEII Fellowship from the Australian Government. 
PJE is supported by the SSimPL programme and the Sydney Institute for Astronomy (SIfA) via ARC grant, DP130100117.
WC and CP acknowledge support of ARC DP130100117.
FF acknowledges financial support from the grants PRIN INAF 2010 "From the dawn of galaxy formation" and PRIN MIUR 2012 "The Intergalactic Medium as a probe of the growth of cosmic structures".
JGB is supported by Spain through the MINECO grant FPA2015-68048, as well as the Consolider-Ingenio 2010 Programme of MICINN under grant PAU
CSD2009-00060 and the Severo Ochoa Programme SEV-2012-0249.
VGP acknowledges support from a European Research Council Starting Grant (DEGAS-259586). This work used the DiRAC Data Centric system at Durham University, operated by the Institute for Computational Cosmology on behalf of the STFC DiRAC HPC Facility (www.dirac.ac.uk). This equipment was funded by BIS National E-infrastructure capital grant ST/K00042X/1, STFC capital grant ST/H008519/1, and STFC DiRAC Operations grant ST/K003267/1 and Durham University. DiRAC is part of the National E-Infrastructure.
The work of BMBH was supported by a Zwicky Prize fellowship and
by Advanced Grant 246797 “GALFORMOD” from the European Research Council.
MH acknowledges financial support from the European Research Council via an Advanced Grant under grant agreement no. 321323 NEOGAL. 
AK is supported by the {\it Ministerio de Econom\'ia y Competitividad} and the {\it Fondo Europeo de Desarrollo Regional} (MINECO/FEDER, UE) in Spain through grants AYA2012-31101 and AYA2015-63810-P as well as the Consolider-Ingenio 2010 Programme of the {\it Spanish Ministerio de Ciencia e Innovaci\'on} (MICINN) under grant MultiDark CSD2009-00064. He also acknowledges support from the {\it Australian Research Council} (ARC) grant DP140100198.
PM has been supported by a FRA2012 grant of the University of Trieste, PRIN2010-2011 (J91J12000450001)	from MIUR, and Consorzio per la Fisica di Trieste.
NDP P was supported by BASAL PFB-06 CATA, and Fondecyt 1150300. Part of the calculations presented here were run using the Geryon cluster at the Center for Astro-Engineering at U. Catolica, which received funding from QUIMAL 130008 and Fondequip AIC-57.  
CP acknowledges support of the Australian Research Council (ARC) through Future Fellowship FT130100041 and Discovery Project DP140100198. 
RSS thanks the Downsbrough family for their generous support. 
PAT acknowledges support from the Science and Technology Facilities Council (grant number ST/L000652/1). 
SKY acknowledges support from the National Research Foundation of Korea (Doyak 2014003730). Numerical simulations were performed using the KISTI supercomputer under the programme of KSC-2013-C3-015.

The authors of the paper contributed in the following ways: AP, RAS and EG lead the study and wrote the paper. AK, JGB and FRP organized the second week of the nIFTy workshop from where this study began. The authors listed in Table \ref{tab:models} performed their galaxy formation models using their code, in particular AB, FJC, AC, SC, DC, GDL, FF, VGP, BH, JL, PM, RAS, RS, CVM, and SY actively ran their models with the assistance of JH, MH, and CS. WC, DC, PJE, CP, and JO assisted with the analysis and data format issues. All authors had the opportunity to proof read and comment on the paper.

\bibliography{aamnem99,biblist}

\appendix

\section{Impact of halo mass definition}\label{sec:mass_comp}

\begin{figure}
\begin{centering}
\includegraphics[scale = .41]{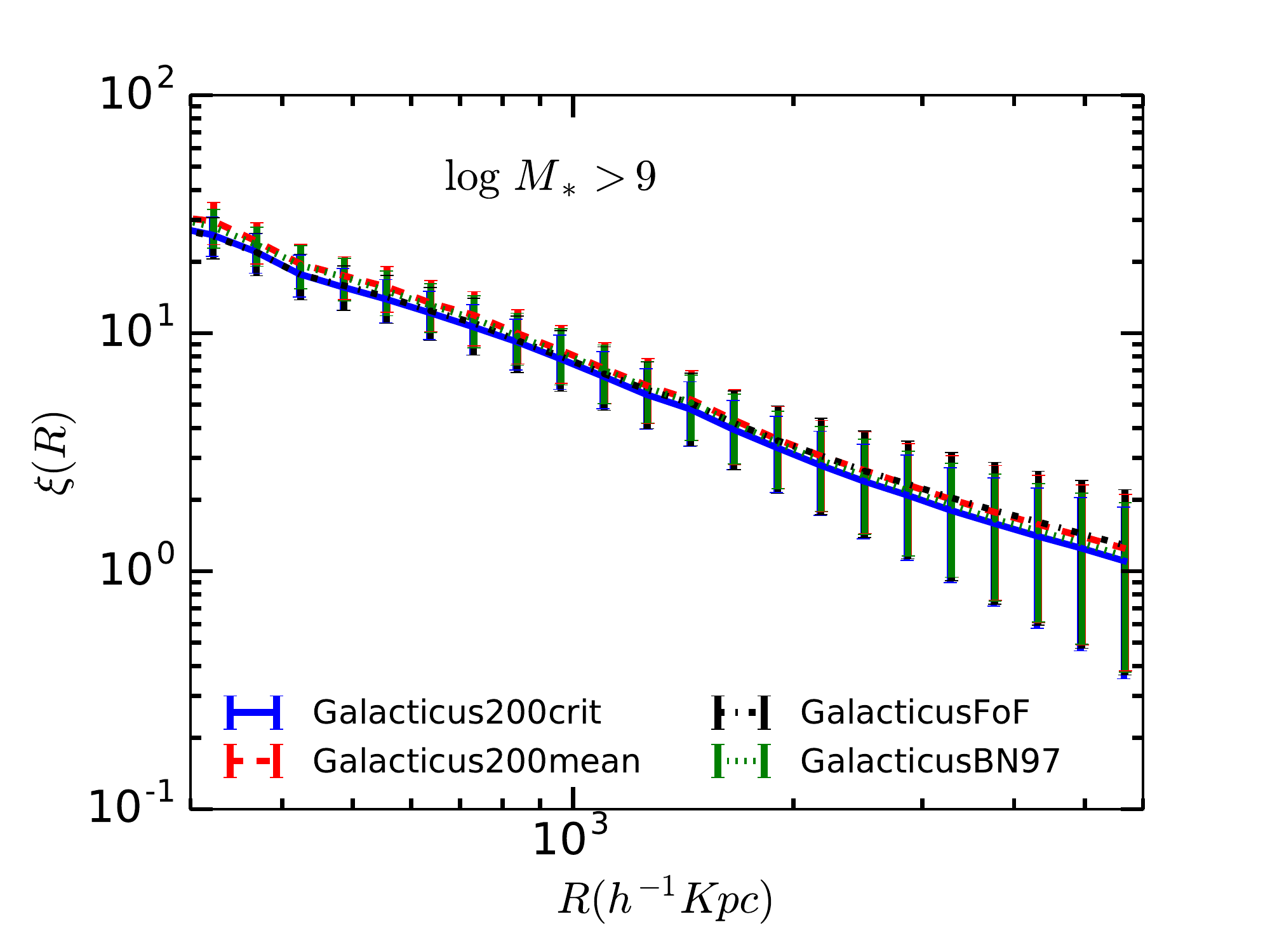}\caption[galaxy 2PCF for \galacticus\ model with different definitions of halo mass]{2PCF for the \galacticus\ model for galaxies with $M* > 10^{10}\Mo$. Each line represents a different halo mass definition used for the model.} 
\label{fig:mass_comp_2pcf}
\par\end{centering}
\end{figure}

Halo mass is one of the properties that galaxy formation models use to determine the population and properties of galaxies. Because of this the models might produce different results if they use different definitions of halo mass. In order to make fair comparisons in our analysis we need to study the dependence on the halo mass definition. 

In Fig. \ref{fig:mass_comp_2pcf} we show  the 2-Point Correlation Function (2PCF)  for different halo mass definitions using the \galacticus\ model for galaxies with $M_* > 10^{10} \Mo$. We only show this model because it used all the different mass definitions and also because we expect the other models to show similar behaviour. Although we only present one model, the different mass definitions reveal only very small changes to the clustering  compared with the differences between the models. Thus, the clustering of the models is not affected by the halo mass definition significantly. This result is also independent of the stellar mass selection used. Hence, the results of our study do not depend on the masses used and we focus on few definitions. In this paper we use $M_{200m}$ (defined as the mass enclosed in a radius within the density is $200$ times the mean density) in all the models which used this mass to obtain a catalogue, and $M_{200c}$ (defined as the mass enclosed in a radius within the density is $200$ times the critical density) or $M_{FOF}$ (defined from the total number of particles belonging to the FOF group) for the other  models, which do not use $M_{200m}$. The results and conclusions of the paper do not depend on the mass definitions used.

\label{lastpage}

\end{document}